\PassOptionsToPackage{expansion=false}{microtype}
\documentclass[sigconf,nonacm]{acmart}
\newcommand{\mytitle}{Comparing Performance of General Context-Free Parsers}
\renewcommand{\mytitle}{An Empirical Comparison of General Context-Free Parsers}

\setcopyright{none}
\copyrightyear{2026}
\acmYear{2026}

\usepackage{hyperref}
\hypersetup{
    colorlinks = true,
    linkcolor = blue,
    anchorcolor = blue,
    citecolor = blue,
    filecolor = blue,
    urlcolor = blue
    }
\usepackage{xspace} 
\usepackage{changepage}   
\usepackage{array}   

\usepackage{cleveref}
\usepackage[inline]{enumitem}
\usepackage{multirow}
\usepackage{comment}
\usepackage{tikz}
\usetikzlibrary{automata, positioning}
\usetikzlibrary{arrows, arrows.meta}
\usetikzlibrary{calc}
\usepackage{pgfplots}
\pgfplotsset{compat=1.18}
\usepackage{textgreek}
\usepackage{float}
\usepackage{caption}
\usepackage{subcaption}
\usepackage{forest}
\usepackage{wasysym} 
\usepackage{proof}
\DeclareCaptionType{Grammar}

\usepackage[most]{tcolorbox}

\newtcolorbox{result}{
  colback=gray!5,           
  colframe=gray!50,         
  boxrule=0.4pt,
  arc=3pt,                  
  left=6pt, right=6pt,
  top=4pt, bottom=4pt,
  center,
  width=1\linewidth,
  fontupper=\small
}

\usepackage{stackengine}
\stackText

\newcounter{todocounter}
\newcommand{\todo}[1]{\marginpar{$|$}\textcolor{red}{\stepcounter{todocounter}\footnote[\thetodocounter]{\textcolor{red}{\textbf{TODO }}\textit{#1}}}}
\newcommand{\done}[1]{\marginpar{$*$}\textcolor{green}{\stepcounter{todocounter}\footnote[\thetodocounter]{\textcolor{black}{\textbf{DONE }}\textit{#1}}}}

\renewcommand{\todo}[1]{}
\renewcommand{\done}[1]{}

\def\nonterm#1{{\color{nontermcolor}\textlangle\textnormal{\emph{#1}}\textrangle}}

\def\expandsto{\(\rightarrow{}\)}

\usepackage{syntax}
\usepackage{tikz-qtree}

\renewcommand{\ulitleft}{\normalfont\ttfamily}
\renewcommand{\litleft}{\bgroup\color{termcolor}`\ulitleft}
\renewcommand{\litright}{\ulitright'\egroup}

\renewcommand{\syntleft}{\bgroup\color{nontermcolor}$\langle$\normalfont\itshape}
\renewcommand{\syntright}{$\rangle$\egroup}

{%
\end{grammar}%
\end{small}%
}

\usepackage{xcolor}
\usepackage{colortbl}
\usepackage{graphicx}
\newcommand{\best}[1]{\cellcolor{blue!25}#1}
\newcommand{\worst}[1]{\cellcolor{orange!30}#1}

\def\|#1|{\textit{#1}} 
\def\<#1>{\texttt{#1}} 

\definecolor{light-gray}{gray}{0.77}

\definecolor{nontermcolor}{rgb}{0.4, 0.05, 0.0} 
\definecolor{absnontermcolor}{rgb}{0.4, 0.5, 0.0} 
\definecolor{evokcolor}{rgb}{0.8, 0.05, 0.1}    
\definecolor{incorrectcolor}{rgb}{0.5, 0.0, 0.13}
\definecolor{incompletecolor}{rgb}{0.0, 0.0, 0.55}
\definecolor{validcolor}{rgb}{0.33, 0.42, 0.18} 
\definecolor{retcolor}{rgb}{0.65, 0.16, 0.16}
\definecolor{ipatterncolor}{rgb}{0.0, 0.5, 0.4} 
\definecolor{termcolor}{rgb}{0.0, 0.05, 0.4}    
\definecolor{regexcolor}{rgb}{0.1, 0.3, 0.1}    

\newcommand{\millisecond}{ms\xspace}

\title{\mytitle}

\author{Huan Vo}
\affiliation{\institution{University of Sydney}\country{Australia}}
\email{hovo3634@uni.sydney.edu.au}

\author{Danushka Liyanage}
\affiliation{\institution{University of Sydney}\country{Australia}}
\email{danushka.liyanage@sydney.edu.au}

\author{Hong Jin Kang}
\affiliation{\institution{University of Sydney}\country{Australia}}
\email{hongjin.kang@sydney.edu.au}

\author{Sasha Rubin}
\affiliation{\institution{University of Sydney}\country{Australia}}
\email{sasha.rubin@sydney.edu.au}

\author{Rahul Gopinath}
\affiliation{\institution{University of Sydney}\country{Australia}}
\email{rahul.gopinath@sydney.edu.au}

\begin{abstract}
Parsing underpins a vast range of software engineering tasks---%
from compilers and static analyzers
to language servers and fuzz testing tools.
Yet most parsers deployed in practice are deterministic (LL or LR),
forcing developers not only to contort their grammars to fit the parser,
but to simplify the very languages they design---%
sacrificing expressiveness for the sake of parseability.
General context-free parsers eliminate this constraint,
but despite decades of algorithmic development,
no rigorous head-to-head comparison exists across the major families of parsing algorithms.

We present the first unified, controlled benchmark
of six generalized parsing algorithms---%
CYK, Valiant, Earley, GLL, RNGLR, and BRNGLR---%
plus deterministic LL(1) and LR(1) baselines,
all implemented in Rust
with shared data structures and parse-tree extraction,
and evaluated across 22~grammars
ranging from simple expressions to full C++ and Java.
Our results show that the cost of generality is lower than widely assumed.
On deterministic grammars, the GLR family incurs only a 3$\times$ median slowdown over LR(1),
with a narrow and predictable variance.
GLR is the clear performance winner among generalized parsers
and a practical default choice for software engineering tools.
\end{abstract}

\begin{document}

\newcommand\sr[1]{}
\newcommand\hj[1]{}

\maketitle

\section{Introduction}

Parsing is a cornerstone of software engineering.
Every tool that reads, analyzes,
or transforms source code such as
compilers, static analyzers, language servers, IDEs,
fuzzers, and program transformation frameworks %
depends on a parser as its front end.
The choice of parsing algorithm, and hence the expressivity of the supported language, has far-reaching consequences
for the flexibility, robustness, and maintainability
of software engineering tooling.

\done{SR: 1. i think you mean "a? the? subset of the deterministic context-free languages". 2. say what $n$ is.}

Despite this centrality, most deployed parsers are limited to
the deterministic subset of context-free languages:
either top-down LL or bottom-up LR.
These algorithms are fast---$O(n)$ in input length---but accept only a restricted subclass of context-free grammars.
In practice, this forces developers to engage in \emph{grammar hacking}~\cite{lammel2001grammar,klint2005toward}:
manually rewriting grammars to eliminate left recursion,
resolve shift-reduce conflicts,
or otherwise contort the structure of the language
\done{SR: "natural language structure" is an odd phrase. it sounds like "natural language" (e.g., English). the point seems to be that this involves changing the language, not just the grammar, in an unnatural way. right? -- updated RG}
to satisfy parser constraints. 

\done{SR: "limited inroads iNto software engineering practice." feels like overselling. Perhaps add a colon after "practice" will fix this --- we explain what we mean further down.}

\done{SR: fit on a business card. cute. but perhaps too informal? sounds like an llm wrote that sentence, which is probably not an ideal look --- Douglas Crockford famously used to distributed his card with JSON grammar printed on one side :), but I take the point.}

\done{SR: "more vulnerable to security exploits."... do you have citations for this? is it [21]? We explain further with Langsec}

\done{"The root cause is
not that formal parsing is hard."... the root cause of what? low adoption? also, in some sense "formal parsing is hard", i.e., not linear...}

The practical consequence is striking.
Despite parsing theory being well-established
since the early decades of computer science~\cite{kegler2023parsing},
it has made limited inroads into software engineering practice.
Faced with the arcana of LL and LR grammar restrictions,
developers routinely abandon formal parsing altogether
and write parsers by hand~\cite{kirschner2022input,eaton2021parser,eaton2021enumerating}.
GitHub alone hosts over 29,800 repositories containing JSON parsers~\cite{github2026search}---%
a format with a simple grammar. 
This proliferation of ad hoc parsers carries real costs:
hand-written parsers are harder to maintain,
more prone to subtle bugs and vulnerabilities.
The Language-theoretic Security community~\cite{langsec}
has long argued that formal, declarative parsers
are a prerequisite for robust and secure input handling,
yet adoption remains low.
The root cause, we argue, is not that formal grammar-based parsing is difficult to apply in practice.
Instead, the issue is that the perceived complexity of \emph{grammar engineering}
for satisfying deterministic parsers has obscured simpler,
more general alternatives~\cite{tratt2011parsing,lammel2001grammar}. 
During \emph{grammar engineering},
it is common to add one more rule, and find oneself with several
shift-reduce conflicts~\cite{vandenbrand2026current}, effectively stopping further progress.
\done{SR: is "generalized parsers" an accepted term? RG Fixed}

\done{SR: "modern hardware..." this probably needs a citation/evidence -- RG: Well known, hence typically skipped in SE}

While general context-free parsers\footnote{
These are also called \emph{generalized parsers}, which we will use interchangeably.} such as CYK \cite{younger1967recognition}, Earley \cite{earley1970an},
GLR \cite{tomita1986efficient}, and GLL \cite{scott2010gll}
accept any context-free grammar without modification, 
they were historically dismissed as impractical
due to their $O(n^3)$ worst-case complexity~\cite{grune2008introduction}.
However, they have seen renewed interest in tools like
Tree-sitter \cite{brunsfeld2024tree}, which uses incremental GLR for syntax highlighting, Parglare~\cite{dejanovic2022parglare}, a complete GLR parser,
and GNU Bison, which now offers a GLR mode.
Other approaches, such as Parsing Expression Grammars~\cite{ford2004parsing},
sidestep the problem 
by using ordered choice---%
but at the cost of redefining language semantics
and complicating grammar composition.

\done{SR: i'm not sure that the previous paragraphs have demonstrated "decades of development across these algorithm families". also,  "families" is used often, and only explained in line 110-, so perhaps some explanation can come earlier? -- RG: SE}

\done{SR: i think for each existing evaluation type of narrowness you need at least once citation, or at least a forward reference to where these are discussed (presumably "Related Work" section later) RG: forward reference}

Yet, despite decades of development across these algorithm families~\cite{kegler2023parsing},
no rigorous head-to-head comparison exists.
As later discussed, existing evaluations are narrow:
they typically cover a single algorithm family,
use different implementation languages,
different data structures,
or different output representations,
making direct comparison impossible.
It remains unclear, in practice,
how much performance is sacrificed by choosing a generalized parser,
or which generalized algorithm is best suited for software engineering tasks.
These open questions leave the misconception of prohibitive cost unchallenged,
and the status quo of proliferating hand-written parsers intact.

We close this gap by implementing six generalized algorithms
across all four major parsing families:
Earley and GLL (online, top-down),
RNGLR and BRNGLR (online, bottom-up GLR variants),
and matrix multiplication based parsers such as
CYK and Valiant (offline, bottom-up)---%
alongside deterministic LL(1) and LR(1) baselines.
All implementations are written in Rust,
a compiled systems language without garbage collection,
using shared grammar representations
and shared data structures.
This controlled environment allows fair,
direct comparison across 22 grammars ranging in complexity
from simple expressions to full C++ and Java.

\noindent We address the following research questions:
\begin{description}[leftmargin=0pt]
\item[RQ1.] How do various families of general-context-free parsing algorithms such as matrix-based, top-down, and bottom-up compare in terms of parsing efficiency?

\done{SR: it is not clear what "parsing efficiency" means, or if it is controversial how to measure it. -- RG will be detailed in later sections.}
\item[RQ2.] How do deterministic parsing algorithms such as LL(1) and LR(1) compare against
      practical general context-free parsing algorithms such as BRNGLR, RNGLR, GLL, and Earley?
\done{SR: you mean, how do they compare on deterministic grammars? RG: Different ways of evaluation; we have det grammars, and non-det grammars of the same family.}
\item[RQ3.] How effective is \emph{grammar-hacking} where a general-context-free grammar is refactored to conform to LR(1) or LL(1) constraints?

\end{description}

Our results show that the GLR family of parsers is the fastest
among the generalized parsers in our study.
On deterministic grammars, the GLR family incurs only a $3\times$ median slowdown
compared to the LR(1) parser.
GLL is slower, with a ${\sim}6\times$ median overhead over LR(1),
while Earley is the slowest, at ${\sim}10\times$ median overhead.

\noindent We make the following contributions:
\begin{itemize}
    \item The first unified, controlled benchmark
    of six generalized parsing algorithms across four algorithm families
    on real-world software engineering grammars.
    \item A detailed comparison of the parser performance
    on refactored grammars recognizing the same language.
    \item Practical guidance for SE tool builders
    on when to use generalized parsing as a default, and when grammar hacking can be counterproductive.
\end{itemize}



\section{Background}
\label{sec:background}
\subsection{Context-Free Grammars and Parsing}
Consider the arithmetic expression \texttt{(1+2)+3}.
The parentheses make the grouping explicit:
add \texttt{2} to \texttt{1} first,
then add \texttt{3} to the result.
Without parentheses, as in \texttt{1+2-3}, the
situation is less clear:
grouping the addition first gives \texttt{(1+2)-3},
while grouping the subtraction first gives \texttt{1+(2-3)},
and these produce different values since subtraction is not associative.
A parser's job is to recover this structural information from a flat string of characters.

The primary output of a parser is a \textit{parse tree} (or \textit{derivation tree}):
a rooted tree whose root represents the whole input,
whose internal nodes represent syntactic categories such as \nonterm{expr} or \nonterm{number},
and whose leaves are the individual input tokens.
\Cref{fig:exprGrammar} shows a grammar for arithmetic expressions,
which we use as our running example throughout this section.

\begin{figure}[h!]
\begin{grammar}
<expr> $\rightarrow$ <expr> `+' <expr> | <expr> `-' <expr> | `(' <expr> `)'
      \alt <number>

<number> $\rightarrow$ <digit> <number> | <digit>

<digit> $\rightarrow$ `0' | `1' | `2' | `3' | `4' | `5' | `6' | `7' | `8' | `9'
\end{grammar}
\caption{An arithmetic expression grammar}
\label{fig:exprGrammar}
\end{figure}

\begin{figure}[h!]
\begin{tikzpicture}[
  level distance=0.6cm,
  sibling distance=0.2cm,
  every tree node/.style={
    font=\small,
    inner sep=2pt,
    text height=1.5ex,
    text depth=0.5ex
  },
  every leaf node/.style={
    font=\small,
    inner sep=1pt,
    level distance=0.4cm,
    text height=,
    text depth=,
  },
]
\Tree[.\nonterm{expr}
  [.\nonterm{expr}
    \texttt{(}
    [.\nonterm{expr}
      [.\nonterm{number} [.\nonterm{digit} \texttt{1} ] ]
      \texttt{+}
      [.\nonterm{number}
        [.\nonterm{digit} \texttt{2} ]
      ]
    ]
    \texttt{)}
  ]
  \texttt{+}
  [.\nonterm{expr}
    [.\nonterm{number}
      [.\nonterm{digit} \texttt{3} ]
    ]
  ]
]
\end{tikzpicture}
\caption{Expression grammar parse tree for \texttt{(1+2)+3}}
\label{fig:parsetree1}
\end{figure}
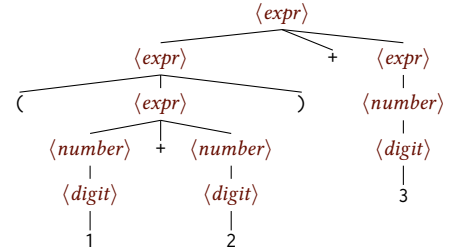

The grammar consists of \textit{production rules},
each of which says that a syntactic category may be formed in a particular way.
Together the rules define every valid string and every valid parse tree in the language.
\Cref{fig:parsetree1} shows the parse tree that the grammar assigns to \texttt{(1+2)+3}:
the parentheses force the left \nonterm{expr} to span \texttt{(1+2)},
and the outer \nonterm{expr} combines it with \texttt{3} via addition.

For a string such as \texttt{1+2-3},
the grammar permits two trees---one for each grouping---making it \textit{ambiguous}.
This property has consequences for parser design,
examined in the next subsection.


Parsing encompasses two related tasks.
\textit{Recognition} determines whether $w \in L(G)$:
does the input conform to the grammar?
\done{we are not measuring complexity wrt $G$, only $w$. i.e., the grammar is fixed, and the parser takes $w$ as input... right? this is called \"data complexity\" --- same as above.}
\textit{Parse-tree construction} goes further and recovers
one or more derivation trees for $w$.
For an unambiguous grammar every string has at most one tree;
for an ambiguous grammar like \Cref{fig:exprGrammar},
multiple trees may exist and must all be accounted for.
Recognition is the primary benchmark criterion;
tree extraction is a separate pass that does not affect the asymptotic cost
(\Cref{sec:methodology}).
\done{SR: i'm not sure what this means: \"recognition alone is sufficient to characterize the algorithmic complexity of a parsing algorithm.\" -- fixed}

We now make these notions precise.
A \textit{Context-Free Grammar} (CFG) is a 4-tuple $G = (N, \Sigma, P, S)$,
where $N$ is a finite set of \textit{non-terminals} (syntactic categories),
$\Sigma$ is a finite set of \textit{terminals} (the alphabet of the input),
$P$ is a finite set of \textit{production rules} of the form $A \to \alpha$
where $A \in N$ and $\alpha \in (N \cup \Sigma)^*$,
and $S \in N$ is the \textit{start symbol}.
A grammar generates strings by \textit{derivation}:
starting from $S$ (here \nonterm{expr}),
non-terminals are repeatedly replaced by production-rule right-hand sides
until only terminals remain,
simultaneously constructing the parse tree.
Expanding the leftmost non-terminal yields a \textit{leftmost derivation};
expanding the rightmost gives a \textit{rightmost derivation}.
The set of all strings derivable from $S$ is the \textit{language} $L(G)$.

\done{SR: In general, recognizing any CFG takes $O(n^3)$ time in the length of the input $n$"... Careful. I would write "It is a classic result that for every CFG, the recognition problem can be solved in cubic time. --- may be too formal for SE?}

It is a classic result that for every CFG, the recognition problem can be solved in cubic time
\cite{younger1967recognition, earley1970an}.
For most practical purposes this is too slow,
motivating the development of restricted grammar classes
that admit linear-time algorithms.

\subsection{Ambiguity and Determinism}

\done{SR: ambiguity was mentioned earlier, so perhaps it shouldn't be mentioned earlier too --- an informal mention followed by a formal definition is standard in SE venues.}

\done{SR: also, i don't think we need the word 'distinct' if we say 'two or more', but it doesn't hurt to leave it as is if you like. -- distinct is useful for emphasis}

A grammar $G$ is \textit{ambiguous} if some string $w \in L(G)$
has two or more distinct parse trees.
The grammar in \Cref{fig:exprGrammar} is itself ambiguous:
\texttt{1+2-3} admits the two derivations, visualized as proof-tree derivations with only \nonterm{expr} shown for clarity:
\[
\infer{\nonterm{expr}}%
{\infer{\nonterm{expr}}{\texttt{1} & \texttt{+} & \texttt{2}} & \texttt{-} & \texttt{3}}
\qquad\text{and}\qquad
\infer{\nonterm{expr}}%
{\texttt{1} & \texttt{+} & \infer{\nonterm{expr}}{\texttt{2} & \texttt{-} & \texttt{3}}}
\]
These correspond to \texttt{(1+2)-3} and \texttt{1+(2-3)},
which yield different semantic values since subtraction is not associative.
\done{SR: the fact that they yield different values is not relevant to whether or not the grammar is ambiguous.
i would remove that statement. --- the semantic observation (different values) is useful in SE context explaining why ambiguity matters in practice;}
\done{SR: also, your trees are suddenly inverted... '.}

The classic \textit{dangling else} illustrates a deeper difficulty:
even a natural, unambiguous-\textit{seeming} language construct
can resist a clean deterministic grammar.
The rule
\begin{grammar}
<stmt> $\rightarrow$ `if' <E> `then' <stmt> 
      \alt `if' <E> `then' <stmt> `else' <stmt>
\end{grammar}

is ambiguous:
in \texttt{if}~$a$~\texttt{then if}~$b$~\texttt{then}~$s_1$~\texttt{else}~$s_2$,
the \texttt{else} can attach to either \texttt{if}.
Most languages resolve this informally---associate
\texttt{else} with the nearest unmatched \texttt{if}, but
the intuitive representation of the grammar does not encode this rule,
and no clean unambiguous rewrite exists without introducing auxiliary non-terminals
that obscure the grammar's intent.
C and Java sidestep the problem in the parser rather than the grammar,
quietly resolving the shift-reduce conflict in favor of shifting.
Similarly, for the expression grammar, an addition rule is intuitively represented as:
\nonterm{E} \expandsto \nonterm{E} `+' \nonterm{E}, and not as \nonterm{E} \expandsto \nonterm{T} `+' \nonterm{F}

\done{SR: in case it is a problem, i would note that the phrasing in the last sentence is something an llm would write. --- fixed}

\subsection{Deterministic Parsers}
\label{sec:deterministicParsers}

Deterministic parsers exploit the LL or LR structure of a grammar
to parse in $O(n)$ time.

\noindent\textbf{LL parsers.}
LL parsers work \textit{top-down},
building the parse tree from the root.
At each step, they predict which production rule to expand
based on the current non-terminal and the next token.

\noindent\textbf{LR parsers.}
LR parsers work \textit{bottom-up},
assembling the parse tree from the leaves upward.
First introduced by Knuth \cite{knuth1965on},
they maintain an explicit state stack
and make decisions---\textit{shift} the next token onto the stack,
or \textit{reduce} the top of the stack to a non-terminal---by
consulting a precomputed parse table.
Because decisions are made after seeing the full right-hand side of a rule,
LR parsers handle left recursion naturally and are strictly more powerful than LL parsers.
However, the parse tables can contain \textit{shift-reduce} or \textit{reduce-reduce} conflicts
when the grammar requires more than $k$ tokens of lookahead,
signalling that the grammar lies outside LR$(k)$.

\subsection{Generalized Parsers}

Generalized parsers accept any CFG,
handling both ambiguity and constructs that fall outside
the deterministic grammar classes. 
Their worst-case runtime is $O(n^\omega)$, where $\omega = 3$ for most parsers.

\begin{table}[h]
\caption{Parsing algorithm families} 
\small
{\centering
\begin{tabular}{|l|l|l|l|l|r|}
\hline
\textbf{Algorithm} & \textbf{Build} & \textbf{Scan} & \textbf{Accepts} & \textbf{$O(\omega)$} & \textbf{LoC} \\
\hline\hline
\multicolumn{6}{|l|}{\textit{Deterministic}} \\[2pt]
\quad LL$(1)$  & Top-down  & Online  & LL$(1)$ only         & $O(n)$        & 254       \\
\quad LR$(1)$  & Bottom-up & Online  & LR$(1)$ only         & $O(n)$        & $923$ \\
\hline
\multicolumn{6}{|l|}{\textit{Generalized --- online, top-down}} \\[2pt]
\quad Earley   & Top-down  & Online  & All CFG              & $O(n^3)$      & 708  \\
\quad GLL      & Top-down  & Online  & All CFG              & $O(n^3)$      & 912  \\
\hline
\multicolumn{6}{|l|}{\textit{Generalized --- online, bottom-up}} \\[2pt]
\quad RNGLR    & Bottom-up & Online  & All CFG              & $O(n^3)$      & $1{,}699$ \\
\quad BRNGLR   & Bottom-up & Online  & All CFG              & $O(n^3)$      & $2{,}015$     \\
\hline
\multicolumn{6}{|l|}{\textit{Generalized --- offline, bottom-up}} \\[2pt]
\quad CYK      & Bottom-up & Offline & All CFG$^\text{CNF}$    & $O(n^3)$      & 143  \\
\quad Valiant$\dagger$   & Bottom-up & Offline & All CFG$^\text{CNF}$    & $O(\frac{n^3}{\log n})$ & 312  \\
\hline
\end{tabular}}\\
\noindent$^\dagger$ Using the method of Four Russians\@.
\label{tab:parserFamilies}
\end{table}

Our work studies algorithms in three families
along two axes: \textit{build direction} (top-down vs.\ bottom-up)
and \textit{scan discipline} (online left-to-right vs.\ offline).
Online algorithms process the input one token at a time, left to right,
and can in principle begin producing output before the input is fully consumed.
Offline algorithms require the complete input string before any work begins,
filling a table over all substrings simultaneously.
\Cref{tab:parserFamilies} summarizes these algorithms 
alongside deterministic baselines from \Cref{sec:deterministicParsers}.
We describe each family in turn.
\begin{table*}[tp]
\small
  \centering
  \caption{%
    Benchmark grammars and their structural properties.
    \emph{Prod.}~= productions;
    \emph{NT}~= non-terminals;
    \emph{Term.}~= terminals;
    \emph{Null.}~= nullable non-terminals;
    \emph{Left Rec.}~= left recursion present;
    \emph{Ambiguous}~= grammar is known to be ambiguous.}
\begin{tabular}{|p{1.5em}|l|l|r|r|r|r|c|c|c|c|}
  \hline
  & \textbf{Grammar} & \textbf{Category} & \textbf{Prod.} & \textbf{NT} & \textbf{Term.} & \textbf{Null.} & \textbf{Left Rec.} & \textbf{Ambiguous} & \textbf{LL(1)} & \textbf{LR(1)} \\
  \hline\hline
  \multirow{4}{*}{\rotatebox{90}{LL(1)}}
   & S-Expr LL-1  & Simple   & 49   & 10  & 39 & 2  & ---         & ---         & \checkmark & \checkmark \\
   & Expr LL-1    & Simple   & 24   & 9   & 16 & 3  & ---         & ---         & \checkmark & \checkmark \\
   & JSON LL-1    & Moderate & 135  & 15  & 94 & 7  & ---         & ---         & \checkmark & \checkmark \\
   & TinyPascal   & Moderate & 68   & 17  & 63 & 9  & ---         & ---         & \checkmark & \checkmark \\
  \hline
  \multirow{5}{*}{\rotatebox{90}{\small LR(1), not LL(1)}}
   & Expr (lr)    & Simple   & 21   & 6   & 16 & 0  & \checkmark & ---         & ---        & \checkmark \\
   & Expr (rr)    & Simple   & 21   & 6   & 16 & 0  & ---         & ---         & ---        & \checkmark \\
   & JSON (rr)    & Moderate & 140  & 20  & 94 & 5  & ---         & ---         & ---        & \checkmark \\
   & JSON (lr)    & Moderate & 140  & 20  & 94 & 5  & \checkmark & ---         & ---        & \checkmark \\
   & TinyC LR-1   & Moderate & 66   & 16  & 54 & 2  & \checkmark & ---         & ---        & \checkmark \\
  \hline
  \multirow{13}{*}{\rotatebox{90}{General context-free}}
   & S-Expression & Simple   & 165  & 20  & 94 & 5  & ---         & \checkmark & ---        & --- \\
   & Bool         & Simple   & 6    & 1   & 10 & 0  & \checkmark & \checkmark & ---        & --- \\
   & Expr (ambig) & Simple   & 18   & 3   & 16 & 0  & \checkmark & \checkmark & ---        & --- \\
   & JSON (ambig) & Moderate & 140  & 20  & 94 & 7  & \checkmark & \checkmark & ---        & --- \\
   & TinyC        & Moderate & 63   & 15  & 45 & 1  & \checkmark & \checkmark & ---        & --- \\
   & ANSI~C       & Complex  & 405  & 85  & 94 & 5  & \checkmark & \checkmark & ---        & --- \\
   & Pascal       & Complex  & 501  & 215 & 95 & 66 & \checkmark & \checkmark & ---        & --- \\
   & Java         & Complex  & 1835 & 532 & 99 & 222 & \checkmark & \checkmark & ---       & --- \\
   & C++          & Complex  & 762  & 161 & 94 & 10 & \checkmark & \checkmark & ---        & --- \\
   & CSS          & Complex  & 582  & 127 & 94 & 56 & \checkmark & \checkmark & ---        & --- \\
   & HTML         & Complex  & 686  & 172 & 94 & 67 & \checkmark & \checkmark & ---        & --- \\
   & Shell        & Complex  & 393  & 32  & 94 & 5  & \checkmark & \checkmark & ---        & --- \\
   & SQL          & Complex  & 623  & 87  & 85 & 4  & \checkmark & \checkmark & ---        & --- \\
  \hline
\end{tabular}

  \label{tab:grammars}
\end{table*}

\section{Methodology}
\label{sec:methodology}

\begin{description}
    \item[CYK]
    CYK (Cocke--Younger \cite{younger1967recognition}--Kasami)
    uses dynamic programming in a bottom-up fashion.
    It fills an $n \times n$ triangular table.
    It runs in consistent $O(n^3)$
    and requires the grammar to be in Chomsky Normal Form (CNF).

    Valiant \cite{valiant1975general} showed that parsing can be reduced
    to boolean matrix multiplication,
    achieving sub-cubic time using fast multiplication algorithms
    such as Strassen's \cite{strassen1969gaussian}.
    In our implementation, we use the Method of Four Russians~\cite{arlazarov1970economical}, which is a subcubic ($O(\frac{n^3}{\log n})$)
    practical matrix multiplication algorithm implemented in the Rust library~\href{https://crates.io/crates/m4ri-rust}{m4ri}\footnote{From here on,
    we use \emph{Valiant} to describe the Valiant algorithm with the Method of Four Russians.}.
    

    \item[Earley]
    The Earley parser \cite{earley1970an} handles any CFG without grammar conversion.
    It operates top-down,
    maintaining at each input position a set of \textit{Earley items}----%
    dotted rules $A \to \alpha \bullet \beta$
    that record how far a rule has been matched and where the match began.
    Its worst-case complexity is $O(n^3)$,
    but it degrades gracefully:
    $O(n^2)$ on unambiguous grammars
    and $O(n)$ on a broad class of deterministic grammars.
    Aycock et al.\ \cite{aycock2002practical} resolved a long-standing issue
    with epsilon rules,
    and Leo \cite{leo1991a} introduced a right-recursion optimization
    that restores linear time on LR$(k)$ grammars
    without requiring lookahead or the LR table.

    \item[Generalized LL]
    GLL, introduced by Scott and Johnstone in 2010 \cite{scott2010gll},
    generalizes LL parsing using a Graph Structured Stack
    \done{SR: 'GSS' is used here but only introduced later.}
    to explore multiple derivations without redundant recomputation.
    It handles left-recursive grammars----%
    LL's central limitation----%
    and can be implemented in a recursive-descent style
    without large precomputed parse tables.

    \item[Generalized LR]
    GLR~\cite{tomita1986efficient}
    extends LR by forking the parse stack at every conflict
    and pursuing all alternatives in parallel.
    A \textit{Graph-Structured Stack} (GSS) shares stack prefixes across branches,
    preventing exponential blow-up in most cases.
    Issues with hidden left recursion and epsilon rules were
    solved by
    Scott and Johnstone through 
    Right-Nulled GLR (RNGLR)~\cite{scott2006right},
    and its binary variant BRNGLR \cite{scott2007brnglr},
    which achieves a strict $O(n^3)$ worst-case bound
    for all grammars.
    Empirical evaluation \cite{johnstone2006evaluating} suggests
    RNGLR and BRNGLR are the best-performing GLR variants in practice.
\end{description}

\noindent\textbf{Benchmark.}
A central challenge in comparing parsers is isolating algorithmic differences
from implementation artifacts:
a parser written in Python will behave very differently from one written in C,
even if both implement the same algorithm.
We therefore implement all parsers in a single unified Rust framework.
Every implementation shares the same grammar representation~\cite{zeller2024fuzzing},
token-stream format, AST structures, and memory management strategy,
and all are compiled with identical optimization settings (\texttt{--release}).
Each parser was implemented directly from its best-known published algorithm
or optimizations without tuning beyond what the publication describes,
so that observed differences reflect the algorithms rather than engineering choices.
\done{but for Earley it says below that you don't just look at the original publication, but also at optimizations...}

To validate correctness,
each implementation was independently reviewed by a second researcher.
For each grammar, we generated test inputs with a grammar-based fuzzer~\cite{zeller2024fuzzing}
and verified that the original string can be reconstructed from the parse tree.
We also evaluated diverse input mutations to verify that every parser
correctly rejects strings outside the grammar.

\done{doesn't it make sense to also test on strings that people actually wrote, e.g., actual C programs? --- Yes, we need to do this; but we didn't so far -- needs fix in future work.}

\noindent\textbf{Parser Implementations.}
We implement the six parsing algorithms mentioned in \Cref{tab:parserFamilies}.
\begin{description}
    \item[CYK] A standard dynamic programming parser.
    \item[Valiant] Matrix-multiplication variant of CYK, using the
      Method of Four Russians~\cite{arlazarov1970economical}.
    \item[Earley] With Aycock–Horspool~\cite{aycock2002practical}
    and Leo's optimization~\cite{leo1991a}.
    \item[GLL] Johnstone's Generalized LL parser~\cite{johnstone2023a}.
    \item[RNGLR] Right-Nulled GLR parser from Economopoulos~\cite{economopoulos2006generalized}.
    \item[BRNGLR] Binary RNGLR variant from Economopoulos~\cite{economopoulos2006generalized}.
\end{description}
We also implement LL(1) and LR(1) parsers for baselines.
LR and GLR parsers use precomputed parse tables.
All parsers operate in scannerless mode.
For Valiant and CYK, the grammars are transformed to CNF prior to parsing. Valiant uses the Method of Four Russians~\cite{arlazarov1970economical}
for boolean matrix multiplication,
via \texttt{m4ri-rust}~\cite{wiggers2019m4ri}.

\noindent\textbf{Grammar Selection.}
We evaluate our parsers against 22~grammars.
The dataset ranges from simple structures (S-Expressions) to
full-scale programming languages (C++, Java). \Cref{tab:grammars} details the size and classification of each grammar.

While generalized parsers consume these grammars directly,
deterministic baselines require refactored versions.
Refactoring includes left-recursion elimination, left-factoring,
and rule inlining to remove indirect recursion.
For TinyC and TinyPascal, keywords were capitalized (\texttt{if}$\to$\texttt{IF})
to simulate a lexer, and dangling-else ambiguity was resolved by rule splitting.
Consequently, the refactored grammar may accept a slightly different language;
this is unavoidable in our scannerless setting.

\noindent\textbf{Inputs.}
Inputs were generated by bounded-depth grammar-based fuzzing:
Only inputs within a defined size is retained;
the upper bound is set so that the fastest parser completes within one second.

\noindent\textbf{Measurement.}
For each (parser, grammar, input) triple, an \textit{iteration} is a complete parse timed from the first token to the final parse-forest node. Each triple runs for at least 10 iterations; if the total time is under 500~\millisecond, iterations are increased up to 20. Short inputs thus run 20 iterations, longer ones 10. We report the \textit{median} iteration time to mitigate transient system noise. Triples with a median parse time exceeding 1 second are excluded from larger input analyses, though individual runs over 1 second are retained. Memory is measured by polling RSS every 1~\millisecond. Experiments are conducted on an Apple M1 with 16~GB RAM, with parsers compiled via \texttt{rustc –release}.

\begin{table}
  \centering
  \caption{Median runtime (\millisecond) of CYK and Valiant. } 
  \small
\begin{tabular}{|l|l|r|r|r|}
  \hline
  \textbf{Grammar} & \textbf{Tokens} & \textbf{CYK (\millisecond)} & \textbf{Valiant (\millisecond)} & \textbf{Earley (\millisecond)} \\
  \hline\hline
  \multirow{3}{*}{S-Expression} & 0–10   & \worst{0.077\,{\footnotesize$\pm$\,0.023}} & \multicolumn{1}{c|}{---} & \best{0.044\,{\footnotesize$\pm$\,0.022}} \\
   & 10–20  & \worst{0.233\,{\footnotesize$\pm$\,0.067}} & \multicolumn{1}{c|}{---} & \best{0.095\,{\footnotesize$\pm$\,0.013}} \\
   & 20–100 & \worst{3.53\,{\footnotesize$\pm$\,4.12}} & \multicolumn{1}{c|}{---} & \best{0.243\,{\footnotesize$\pm$\,0.080}} \\
  \hline
  \multirow{3}{*}{Expr (lr)} & 0–10   & \best{0.018\,{\footnotesize$\pm$\,0.006}} & \worst{8.78\,{\footnotesize$\pm$\,6.06}} & 0.021\,{\footnotesize$\pm$\,0.004} \\
   & 10–20  & 0.083\,{\footnotesize$\pm$\,0.030} & \worst{150\,{\footnotesize$\pm$\,72}} & \best{0.050\,{\footnotesize$\pm$\,0.008}} \\
   & 20–100 & 2.41\,{\footnotesize$\pm$\,2.24} & \worst{796\,{\footnotesize$\pm$\,380}} & \best{0.190\,{\footnotesize$\pm$\,0.067}} \\
  \hline
  \multirow{3}{*}{Bool} & 0–10   & 0.006\,{\footnotesize$\pm$\,0.002} & \worst{2.16\,{\footnotesize$\pm$\,5.82}} & \best{0.004\,{\footnotesize$\pm$\,0.002}} \\
   & 10–20  & 0.025\,{\footnotesize$\pm$\,0.011} & \worst{80.0\,{\footnotesize$\pm$\,60.8}} & \best{0.015\,{\footnotesize$\pm$\,0.006}} \\
   & 20–100 & 0.530\,{\footnotesize$\pm$\,1.141} & \worst{248\,{\footnotesize$\pm$\,376}} & \best{0.140\,{\footnotesize$\pm$\,0.451}} \\
  \hline
  \multirow{3}{*}{Expr (ambig)} & 0–10   & \best{0.008\,{\footnotesize$\pm$\,0.003}} & \worst{1.53\,{\footnotesize$\pm$\,4.16}} & 0.013\,{\footnotesize$\pm$\,0.006} \\
   & 10–20  & \best{0.026} & \worst{26.5} & 0.032 \\
   & 20–100 & 2.26\,{\footnotesize$\pm$\,1.31} & \worst{3306\,{\footnotesize$\pm$\,3414}} & \best{1.17\,{\footnotesize$\pm$\,3.18}} \\
  \hline
  \multirow{3}{*}{Expr (rr)} & 0–10   & \best{0.014\,{\footnotesize$\pm$\,0.007}} & \worst{2.14\,{\footnotesize$\pm$\,6.07}} & 0.015\,{\footnotesize$\pm$\,0.006} \\
   & 10–20  & 0.045 & \worst{39.2} & \best{0.033} \\
   & 20–100 & 4.24\,{\footnotesize$\pm$\,2.34} & \worst{1330} & \best{0.235\,{\footnotesize$\pm$\,0.068}} \\
  \hline
\end{tabular}

  Blue indicates best values, and orange worst values.
  \label{tab:cykValiant}
\end{table}

\begin{table*}[tp]
  \centering
  \vspace{-0.4cm}
  \caption{%
    Median runtime (\millisecond) by grammar and input-size bucket. Blue is best, orange worst.}
  \vspace{-0.2cm}
\begin{tabular}{|p{1.5em}|l|l|r|r|r|r|r|r|}
  \hline
  & \multirow{2}{*}{\textbf{Grammar}} & \multirow{2}{*}{\textbf{Tokens}}
    & \multicolumn{2}{c|}{\textbf{Deterministic (\millisecond)}}
    & \multicolumn{4}{c|}{\textbf{Generalised (\millisecond)}} \\
  & & & \textbf{LL(1)} & \textbf{LR(1)} & \textbf{Earley} & \textbf{GLL} & \textbf{RNGLR} & \textbf{BRNGLR} \\
  \hline\hline
  \multirow{12}{*}{\rotatebox{90}{LL(1)}}
   & \multirow{3}{*}{TinyPascal} & 0–5k & \best{0.059\,{\footnotesize$\pm$\,0.379}} & 0.083\,{\footnotesize$\pm$\,0.477} & \worst{1.09\,{\footnotesize$\pm$\,8.50}} & 0.311\,{\footnotesize$\pm$\,2.042} & 0.227\,{\footnotesize$\pm$\,1.386} & 0.237\,{\footnotesize$\pm$\,1.398} \\
   & & 5k–15k & \best{2.16\,{\footnotesize$\pm$\,0.68}} & 2.75\,{\footnotesize$\pm$\,0.86} & \worst{57.8\,{\footnotesize$\pm$\,21.1}} & 13.6\,{\footnotesize$\pm$\,5.1} & 8.13\,{\footnotesize$\pm$\,2.73} & 8.73\,{\footnotesize$\pm$\,2.84} \\
   & & 15k–30k & \best{5.33\,{\footnotesize$\pm$\,1.01}} & 6.68\,{\footnotesize$\pm$\,1.25} & \worst{157\,{\footnotesize$\pm$\,31}} & 39.7\,{\footnotesize$\pm$\,8.3} & 21.1\,{\footnotesize$\pm$\,4.6} & 23.8\,{\footnotesize$\pm$\,5.0} \\
   & \multirow{3}{*}{S-Expr LL-1} & 0–5k & \best{0.780\,{\footnotesize$\pm$\,0.392}} & 0.907\,{\footnotesize$\pm$\,0.462} & \worst{15.5\,{\footnotesize$\pm$\,9.5}} & 3.62\,{\footnotesize$\pm$\,1.91} & 2.63\,{\footnotesize$\pm$\,1.31} & 2.37\,{\footnotesize$\pm$\,1.18} \\
   & & 5k–15k & \best{2.98\,{\footnotesize$\pm$\,0.62}} & 3.53\,{\footnotesize$\pm$\,0.72} & \worst{85.2\,{\footnotesize$\pm$\,21.0}} & 16.6\,{\footnotesize$\pm$\,3.8} & 10.2\,{\footnotesize$\pm$\,2.2} & 9.18\,{\footnotesize$\pm$\,1.99} \\
   & & 15k–30k & \best{7.17} & 8.19 & \worst{213} & 45.6 & 26.4 & 24.0 \\
   & \multirow{3}{*}{Expr LL-1} & 0–5k & \best{0.028\,{\footnotesize$\pm$\,0.415}} & 0.037\,{\footnotesize$\pm$\,0.524} & \worst{0.353\,{\footnotesize$\pm$\,5.480}} & 0.150\,{\footnotesize$\pm$\,2.181} & 0.104\,{\footnotesize$\pm$\,1.426} & 0.110\,{\footnotesize$\pm$\,1.448} \\
   & & 5k–15k & \best{2.78\,{\footnotesize$\pm$\,0.96}} & 3.53\,{\footnotesize$\pm$\,1.23} & \worst{40.9\,{\footnotesize$\pm$\,17.3}} & 17.1\,{\footnotesize$\pm$\,7.4} & 9.55\,{\footnotesize$\pm$\,3.35} & 9.81\,{\footnotesize$\pm$\,3.43} \\
   & & 15k–30k & \best{6.03\,{\footnotesize$\pm$\,1.29}} & 7.65\,{\footnotesize$\pm$\,1.59} & \worst{100\,{\footnotesize$\pm$\,24}} & 43.1\,{\footnotesize$\pm$\,11.2} & 22.2\,{\footnotesize$\pm$\,5.2} & 22.9\,{\footnotesize$\pm$\,5.2} \\
   & \multirow{3}{*}{JSON LL-1} & 0–5k & \best{0.212\,{\footnotesize$\pm$\,0.591}} & 0.234\,{\footnotesize$\pm$\,0.656} & \worst{5.07\,{\footnotesize$\pm$\,25.78}} & 0.991\,{\footnotesize$\pm$\,2.979} & 0.695\,{\footnotesize$\pm$\,3.356} & 0.641\,{\footnotesize$\pm$\,3.182} \\
   & & 5k–15k & \best{2.84\,{\footnotesize$\pm$\,1.12}} & 3.17\,{\footnotesize$\pm$\,1.14} & \worst{132\,{\footnotesize$\pm$\,53}} & 15.4\,{\footnotesize$\pm$\,7.2} & 25.6\,{\footnotesize$\pm$\,20.0} & 24.5\,{\footnotesize$\pm$\,19.6} \\
   & & 15k–30k & 7.79\,{\footnotesize$\pm$\,2.50} & \best{7.67\,{\footnotesize$\pm$\,1.72}} & \worst{384\,{\footnotesize$\pm$\,105}} & 46.9\,{\footnotesize$\pm$\,11.7} & 115\,{\footnotesize$\pm$\,77} & 112\,{\footnotesize$\pm$\,76} \\
  \hline
  \multirow{13}{*}{\rotatebox{90}{LR(1), not LL(1)}}
   & \multirow{3}{*}{JSON (rr)} & 0–5k & --- & \best{0.170\,{\footnotesize$\pm$\,0.463}} & \worst{4.56\,{\footnotesize$\pm$\,23.75}} & 0.803\,{\footnotesize$\pm$\,2.163} & 0.517\,{\footnotesize$\pm$\,2.090} & 0.466\,{\footnotesize$\pm$\,1.931} \\
   & & 5k–15k & --- & \best{2.24\,{\footnotesize$\pm$\,0.78}} & \worst{125\,{\footnotesize$\pm$\,48}} & 11.3\,{\footnotesize$\pm$\,4.8} & 15.1\,{\footnotesize$\pm$\,10.4} & 14.2\,{\footnotesize$\pm$\,10.2} \\
   & & 15k–30k & --- & \best{5.39\,{\footnotesize$\pm$\,1.19}} & \worst{342\,{\footnotesize$\pm$\,98}} & 32.8\,{\footnotesize$\pm$\,8.6} & 63.6\,{\footnotesize$\pm$\,39.0} & 61.1\,{\footnotesize$\pm$\,38.7} \\
   & \multirow{3}{*}{JSON (lr)} & 0–5k & --- & \best{0.172\,{\footnotesize$\pm$\,0.469}} & \worst{4.13\,{\footnotesize$\pm$\,23.12}} & 1.22\,{\footnotesize$\pm$\,13.87} & 0.505\,{\footnotesize$\pm$\,1.337} & 0.454\,{\footnotesize$\pm$\,1.181} \\
   & & 5k–15k & --- & \best{2.27\,{\footnotesize$\pm$\,0.80}} & \worst{131\,{\footnotesize$\pm$\,57}} & 86.2\,{\footnotesize$\pm$\,65.0} & 6.49\,{\footnotesize$\pm$\,2.52} & 5.71\,{\footnotesize$\pm$\,2.25} \\
   & & 15k–30k & --- & \best{5.49\,{\footnotesize$\pm$\,1.22}} & \worst{393\,{\footnotesize$\pm$\,145}} & 346\,{\footnotesize$\pm$\,204} & 17.4\,{\footnotesize$\pm$\,4.8} & 15.5\,{\footnotesize$\pm$\,4.4} \\
   & \multirow{3}{*}{Expr (lr)} & 0–5k & --- & \best{0.030\,{\footnotesize$\pm$\,0.424}} & \worst{0.266\,{\footnotesize$\pm$\,4.088}} & 0.197\,{\footnotesize$\pm$\,3.041} & 0.085\,{\footnotesize$\pm$\,1.218} & 0.090\,{\footnotesize$\pm$\,1.309} \\
   & & 5k–15k & --- & \best{2.91\,{\footnotesize$\pm$\,0.98}} & \worst{30.9\,{\footnotesize$\pm$\,12.7}} & 24.1\,{\footnotesize$\pm$\,10.6} & 8.08\,{\footnotesize$\pm$\,2.94} & 8.77\,{\footnotesize$\pm$\,3.13} \\
   & & 15k–30k & --- & \best{6.14\,{\footnotesize$\pm$\,1.28}} & \worst{77.3\,{\footnotesize$\pm$\,17.8}} & 60.0\,{\footnotesize$\pm$\,14.2} & 18.6\,{\footnotesize$\pm$\,4.6} & 21.8\,{\footnotesize$\pm$\,4.8} \\
   & \multirow{1}{*}{Expr (rr)} & 0–5k & --- & \best{0.089\,{\footnotesize$\pm$\,0.050}} & \worst{0.810\,{\footnotesize$\pm$\,0.488}} & 0.603\,{\footnotesize$\pm$\,0.345} & 0.246\,{\footnotesize$\pm$\,0.140} & 0.259\,{\footnotesize$\pm$\,0.146} \\
   & \multirow{3}{*}{TinyC LR-1} & 0–5k & --- & \best{0.409\,{\footnotesize$\pm$\,0.341}} & \worst{5.10\,{\footnotesize$\pm$\,4.96}} & 2.85\,{\footnotesize$\pm$\,2.56} & 1.22\,{\footnotesize$\pm$\,1.02} & 1.29\,{\footnotesize$\pm$\,1.07} \\
   & & 5k–15k & --- & \best{2.19\,{\footnotesize$\pm$\,0.65}} & \worst{35.7\,{\footnotesize$\pm$\,14.3}} & 18.4\,{\footnotesize$\pm$\,6.6} & 6.49\,{\footnotesize$\pm$\,2.01} & 6.90\,{\footnotesize$\pm$\,2.15} \\
   & & 15k–30k & --- & \best{4.84\,{\footnotesize$\pm$\,1.29}} & \worst{95.6\,{\footnotesize$\pm$\,29.2}} & 47.3\,{\footnotesize$\pm$\,14.2} & 14.8\,{\footnotesize$\pm$\,5.0} & 16.1\,{\footnotesize$\pm$\,5.5} \\
  \hline
  \multirow{24}{*}{\rotatebox{90}{General context-free}}
   & \multirow{3}{*}{S-Expression} & 0–5k & --- & --- & \worst{5.13\,{\footnotesize$\pm$\,7.59}} & 1.62\,{\footnotesize$\pm$\,1.97} & \best{0.917\,{\footnotesize$\pm$\,1.109}} & 0.992\,{\footnotesize$\pm$\,1.199} \\
   & & 5k–15k & --- & --- & \worst{61.0\,{\footnotesize$\pm$\,26.1}} & 14.6\,{\footnotesize$\pm$\,6.4} & \best{7.20\,{\footnotesize$\pm$\,2.77}} & 7.69\,{\footnotesize$\pm$\,3.01} \\
   & & 15k–30k & --- & --- & \worst{170\,{\footnotesize$\pm$\,43}} & 43.6\,{\footnotesize$\pm$\,10.8} & \best{19.6\,{\footnotesize$\pm$\,5.5}} & 22.2\,{\footnotesize$\pm$\,5.7} \\
   & \multirow{3}{*}{TinyC} & 0–5k & --- & --- & \worst{0.257\,{\footnotesize$\pm$\,4.439}} & 0.149\,{\footnotesize$\pm$\,2.330} & \best{0.065\,{\footnotesize$\pm$\,0.941}} & 0.072\,{\footnotesize$\pm$\,0.990} \\
   & & 5k–15k & --- & --- & \worst{50.5\,{\footnotesize$\pm$\,18.0}} & 23.7\,{\footnotesize$\pm$\,8.9} & \best{8.22\,{\footnotesize$\pm$\,2.62}} & 8.94\,{\footnotesize$\pm$\,2.77} \\
   & & 15k–30k & --- & --- & \worst{125\,{\footnotesize$\pm$\,28}} & 61.2\,{\footnotesize$\pm$\,13.1} & \best{19.7\,{\footnotesize$\pm$\,4.8}} & 22.8\,{\footnotesize$\pm$\,5.1} \\
   & \multirow{1}{*}{Bool} & 0–5k & --- & --- & 0.016\,{\footnotesize$\pm$\,0.874} & \worst{0.021\,{\footnotesize$\pm$\,0.269}} & \best{0.009\,{\footnotesize$\pm$\,0.261}} & 0.011\,{\footnotesize$\pm$\,0.230} \\
   & \multirow{1}{*}{Expr (ambig)} & 0–5k & --- & --- & 15.3\,{\footnotesize$\pm$\,247.5} & \worst{16.6\,{\footnotesize$\pm$\,49.6}} & 7.86\,{\footnotesize$\pm$\,25.74} & \best{6.61\,{\footnotesize$\pm$\,19.52}} \\
   & \multirow{3}{*}{JSON (ambig)} & 0–5k & --- & --- & \worst{4.80\,{\footnotesize$\pm$\,23.89}} & 0.829\,{\footnotesize$\pm$\,2.192} & 0.524\,{\footnotesize$\pm$\,2.087} & \best{0.472\,{\footnotesize$\pm$\,1.937}} \\
   & & 5k–15k & --- & --- & \worst{122\,{\footnotesize$\pm$\,46}} & \best{11.3\,{\footnotesize$\pm$\,4.9}} & 14.9\,{\footnotesize$\pm$\,10.5} & 14.2\,{\footnotesize$\pm$\,10.2} \\
   & & 15k–30k & --- & --- & \worst{356\,{\footnotesize$\pm$\,100}} & \best{34.4\,{\footnotesize$\pm$\,9.0}} & 64.0\,{\footnotesize$\pm$\,39.4} & 61.3\,{\footnotesize$\pm$\,39.0} \\
   & \multirow{2}{*}{ANSI C} & 0–5k & --- & --- & \worst{15.5\,{\footnotesize$\pm$\,63.0}} & 8.19\,{\footnotesize$\pm$\,32.99} & 5.33\,{\footnotesize$\pm$\,14.02} & \best{4.43\,{\footnotesize$\pm$\,12.66}} \\
   & & 5k–15k & --- & --- & \worst{276\,{\footnotesize$\pm$\,70}} & 151\,{\footnotesize$\pm$\,42} & 60.3\,{\footnotesize$\pm$\,14.1} & \best{54.9\,{\footnotesize$\pm$\,12.8}} \\
   & \multirow{3}{*}{Pascal} & 0–5k & --- & --- & \worst{11.2\,{\footnotesize$\pm$\,32.3}} & 4.64\,{\footnotesize$\pm$\,14.24} & 3.33\,{\footnotesize$\pm$\,7.90} & \best{3.25\,{\footnotesize$\pm$\,7.76}} \\
   & & 5k–15k & --- & --- & \worst{160\,{\footnotesize$\pm$\,62}} & 74.1\,{\footnotesize$\pm$\,30.4} & \best{38.4\,{\footnotesize$\pm$\,14.6}} & 38.8\,{\footnotesize$\pm$\,14.4} \\
   & & 15k–30k & --- & --- & \worst{460\,{\footnotesize$\pm$\,64}} & 214\,{\footnotesize$\pm$\,26} & 105\,{\footnotesize$\pm$\,12} & \best{103\,{\footnotesize$\pm$\,13}} \\
   & \multirow{2}{*}{Java} & 0–5k & --- & --- & \worst{7.75\,{\footnotesize$\pm$\,52.69}} & 3.62\,{\footnotesize$\pm$\,30.84} & 1.29\,{\footnotesize$\pm$\,4.93} & \best{1.25\,{\footnotesize$\pm$\,4.78}} \\
   & & 5k–15k & --- & --- & \worst{299\,{\footnotesize$\pm$\,73}} & 198\,{\footnotesize$\pm$\,46} & 29.4\,{\footnotesize$\pm$\,6.7} & \best{28.0\,{\footnotesize$\pm$\,6.4}} \\
   & \multirow{2}{*}{C++} & 0–5k & --- & --- & \worst{91.4\,{\footnotesize$\pm$\,148.5}} & 58.6\,{\footnotesize$\pm$\,85.4} & 15.0\,{\footnotesize$\pm$\,24.3} & \best{13.9\,{\footnotesize$\pm$\,21.7}} \\
   & & 5k–15k & --- & --- & \worst{1151} & 365\,{\footnotesize$\pm$\,96} & 128\,{\footnotesize$\pm$\,57} & \best{112\,{\footnotesize$\pm$\,46}} \\
   & \multirow{1}{*}{CSS} & 0–5k & --- & --- & \worst{4.30\,{\footnotesize$\pm$\,45.95}} & 1.15\,{\footnotesize$\pm$\,6.43} & 1.22\,{\footnotesize$\pm$\,10.36} & \best{0.993\,{\footnotesize$\pm$\,8.003}} \\
   & \multirow{1}{*}{HTML} & 0–5k & --- & --- & \worst{2.57\,{\footnotesize$\pm$\,15.62}} & 1.08\,{\footnotesize$\pm$\,10.29} & 0.903\,{\footnotesize$\pm$\,34.885} & \best{0.732\,{\footnotesize$\pm$\,25.722}} \\
   & \multirow{1}{*}{Shell} & 0–5k & --- & --- & 3.99\,{\footnotesize$\pm$\,8.75} & \worst{8.27\,{\footnotesize$\pm$\,47.32}} & 2.66\,{\footnotesize$\pm$\,16.58} & \best{2.23\,{\footnotesize$\pm$\,14.08}} \\
   & \multirow{1}{*}{SQL} & 0–5k & --- & --- & 0.068\,{\footnotesize$\pm$\,1.694} & \worst{0.078\,{\footnotesize$\pm$\,49.789}} & \best{0.028\,{\footnotesize$\pm$\,0.459}} & 0.036\,{\footnotesize$\pm$\,0.571} \\
  \hline
\end{tabular}

  \vspace{-0.3cm}
  \label{tab:runtimeBucket}
\end{table*}

\begin{table*}[tp]
  \centering
  \vspace{-0.4cm}
  \caption{%
    Median peak memory (MB) by grammar and input-size bucket. Blue is best, orange worst.}
  \vspace{-0.2cm}
\begin{tabular}{|p{1.5em}|l|l|r|r|r|r|r|r|}
  \hline
  & \multirow{2}{*}{\textbf{Grammar}} & \multirow{2}{*}{\textbf{Tokens}}
    & \multicolumn{2}{c|}{\textbf{Deterministic (MB)}}
    & \multicolumn{4}{c|}{\textbf{Generalised (MB)}} \\
  & & & \textbf{LL(1)} & \textbf{LR(1)} & \textbf{Earley} & \textbf{GLL} & \textbf{RNGLR} & \textbf{BRNGLR} \\
  \hline\hline
  \multirow{12}{*}{\rotatebox{90}{LL(1)}}
   & \multirow{3}{*}{TinyPascal} & 0–5k & 0.016\,{\footnotesize$\pm$\,0.000} & 0.016\,{\footnotesize$\pm$\,0.000} & 0.016\,{\footnotesize$\pm$\,0.008} & 0.016\,{\footnotesize$\pm$\,0.000} & 0.016\,{\footnotesize$\pm$\,0.000} & 0.016\,{\footnotesize$\pm$\,0.000} \\
   & & 5k–15k & 0.016\,{\footnotesize$\pm$\,0.000} & 0.016\,{\footnotesize$\pm$\,0.000} & 0.016\,{\footnotesize$\pm$\,0.000} & 0.016\,{\footnotesize$\pm$\,0.089} & 0.016\,{\footnotesize$\pm$\,0.000} & 0.016\,{\footnotesize$\pm$\,0.000} \\
   & & 15k–30k & \best{0.016\,{\footnotesize$\pm$\,0.000}} & \best{0.016\,{\footnotesize$\pm$\,0.000}} & \best{0.016\,{\footnotesize$\pm$\,0.000}} & \worst{2.38\,{\footnotesize$\pm$\,1.07}} & \best{0.016\,{\footnotesize$\pm$\,0.807}} & \best{0.016\,{\footnotesize$\pm$\,0.928}} \\
   & \multirow{3}{*}{S-Expr LL-1} & 0–5k & 0.016\,{\footnotesize$\pm$\,0.000} & 0.016\,{\footnotesize$\pm$\,0.000} & 0.016\,{\footnotesize$\pm$\,0.000} & 0.016\,{\footnotesize$\pm$\,0.002} & 0.016\,{\footnotesize$\pm$\,0.000} & 0.016\,{\footnotesize$\pm$\,0.000} \\
   & & 5k–15k & \best{0.016\,{\footnotesize$\pm$\,0.000}} & \best{0.016\,{\footnotesize$\pm$\,0.000}} & \best{0.016\,{\footnotesize$\pm$\,0.000}} & \worst{0.531\,{\footnotesize$\pm$\,0.088}} & 0.109\,{\footnotesize$\pm$\,0.133} & \best{0.016\,{\footnotesize$\pm$\,0.000}} \\
   & & 15k–30k & \best{0.016} & \best{0.016} & 4.16 & \worst{4.41} & 0.344 & \best{0.016} \\
   & \multirow{3}{*}{Expr LL-1} & 0–5k & 0.016\,{\footnotesize$\pm$\,0.013} & 0.016\,{\footnotesize$\pm$\,0.000} & 0.016\,{\footnotesize$\pm$\,0.006} & 0.016\,{\footnotesize$\pm$\,0.003} & 0.016\,{\footnotesize$\pm$\,0.002} & 0.016\,{\footnotesize$\pm$\,0.002} \\
   & & 5k–15k & 0.016\,{\footnotesize$\pm$\,0.000} & 0.016\,{\footnotesize$\pm$\,0.002} & 0.016\,{\footnotesize$\pm$\,0.004} & 0.062\,{\footnotesize$\pm$\,1.018} & 0.016\,{\footnotesize$\pm$\,0.000} & 0.016\,{\footnotesize$\pm$\,0.033} \\
   & & 15k–30k & \best{0.016\,{\footnotesize$\pm$\,0.000}} & \best{0.016\,{\footnotesize$\pm$\,0.000}} & \best{0.016\,{\footnotesize$\pm$\,0.004}} & \worst{2.42\,{\footnotesize$\pm$\,1.24}} & \best{0.016\,{\footnotesize$\pm$\,0.553}} & \best{0.016\,{\footnotesize$\pm$\,0.829}} \\
   & \multirow{3}{*}{JSON LL-1} & 0–5k & 0.016\,{\footnotesize$\pm$\,0.000} & 0.016\,{\footnotesize$\pm$\,0.000} & 0.016\,{\footnotesize$\pm$\,0.003} & 0.016\,{\footnotesize$\pm$\,0.023} & 0.016\,{\footnotesize$\pm$\,0.000} & 0.016\,{\footnotesize$\pm$\,0.035} \\
   & & 5k–15k & 0.016\,{\footnotesize$\pm$\,0.003} & 0.016\,{\footnotesize$\pm$\,0.000} & 0.016\,{\footnotesize$\pm$\,0.000} & 0.016\,{\footnotesize$\pm$\,0.653} & 0.016\,{\footnotesize$\pm$\,0.000} & 0.016\,{\footnotesize$\pm$\,0.000} \\
   & & 15k–30k & 0.016\,{\footnotesize$\pm$\,0.003} & 0.016\,{\footnotesize$\pm$\,0.000} & 0.016\,{\footnotesize$\pm$\,0.024} & 0.297\,{\footnotesize$\pm$\,1.383} & 0.016\,{\footnotesize$\pm$\,0.448} & 0.016\,{\footnotesize$\pm$\,0.304} \\
  \hline
  \multirow{13}{*}{\rotatebox{90}{LR(1), not LL(1)}}
   & \multirow{3}{*}{JSON (rr)} & 0–5k & --- & 0.016\,{\footnotesize$\pm$\,0.000} & 0.016\,{\footnotesize$\pm$\,0.031} & 0.016\,{\footnotesize$\pm$\,0.134} & 0.016\,{\footnotesize$\pm$\,0.004} & 0.016\,{\footnotesize$\pm$\,0.002} \\
   & & 5k–15k & --- & \best{0.016\,{\footnotesize$\pm$\,0.002}} & 0.062\,{\footnotesize$\pm$\,0.038} & \best{0.016\,{\footnotesize$\pm$\,0.054}} & \best{0.016\,{\footnotesize$\pm$\,0.005}} & \best{0.016\,{\footnotesize$\pm$\,0.003}} \\
   & & 15k–30k & --- & \best{0.016\,{\footnotesize$\pm$\,0.010}} & 0.062\,{\footnotesize$\pm$\,0.035} & \worst{0.828\,{\footnotesize$\pm$\,1.359}} & \best{0.016\,{\footnotesize$\pm$\,0.487}} & \best{0.016\,{\footnotesize$\pm$\,0.703}} \\
   & \multirow{3}{*}{JSON (lr)} & 0–5k & --- & 0.016\,{\footnotesize$\pm$\,0.000} & 0.016\,{\footnotesize$\pm$\,0.000} & 0.016\,{\footnotesize$\pm$\,0.007} & 0.016\,{\footnotesize$\pm$\,0.000} & 0.016\,{\footnotesize$\pm$\,0.000} \\
   & & 5k–15k & --- & 0.016\,{\footnotesize$\pm$\,0.000} & 0.016\,{\footnotesize$\pm$\,0.002} & 0.016\,{\footnotesize$\pm$\,0.087} & 0.016\,{\footnotesize$\pm$\,0.000} & 0.016\,{\footnotesize$\pm$\,0.000} \\
   & & 15k–30k & --- & 0.016\,{\footnotesize$\pm$\,0.000} & 0.016\,{\footnotesize$\pm$\,0.198} & 0.016\,{\footnotesize$\pm$\,0.000} & 0.016\,{\footnotesize$\pm$\,0.560} & 0.016\,{\footnotesize$\pm$\,0.396} \\
   & \multirow{3}{*}{Expr (lr)} & 0–5k & --- & 0.016\,{\footnotesize$\pm$\,0.012} & 0.016\,{\footnotesize$\pm$\,0.036} & 0.016\,{\footnotesize$\pm$\,0.005} & 0.016\,{\footnotesize$\pm$\,0.002} & 0.016\,{\footnotesize$\pm$\,0.000} \\
   & & 5k–15k & --- & 0.016\,{\footnotesize$\pm$\,0.000} & 0.016\,{\footnotesize$\pm$\,0.003} & 0.016\,{\footnotesize$\pm$\,1.225} & 0.016\,{\footnotesize$\pm$\,0.005} & 0.016\,{\footnotesize$\pm$\,0.005} \\
   & & 15k–30k & --- & \best{0.016\,{\footnotesize$\pm$\,0.000}} & \best{0.016\,{\footnotesize$\pm$\,0.005}} & \worst{2.44\,{\footnotesize$\pm$\,1.69}} & \best{0.016\,{\footnotesize$\pm$\,0.769}} & 0.031\,{\footnotesize$\pm$\,1.018} \\
   & \multirow{1}{*}{Expr (rr)} & 0–5k & --- & 0.016\,{\footnotesize$\pm$\,0.000} & 0.016\,{\footnotesize$\pm$\,0.006} & 0.016\,{\footnotesize$\pm$\,0.009} & 0.016\,{\footnotesize$\pm$\,0.002} & 0.016\,{\footnotesize$\pm$\,0.000} \\
   & \multirow{3}{*}{TinyC LR-1} & 0–5k & --- & 0.016\,{\footnotesize$\pm$\,0.000} & 0.016\,{\footnotesize$\pm$\,0.008} & 0.016\,{\footnotesize$\pm$\,0.002} & 0.016\,{\footnotesize$\pm$\,0.000} & 0.016\,{\footnotesize$\pm$\,0.000} \\
   & & 5k–15k & --- & 0.016\,{\footnotesize$\pm$\,0.000} & 0.016\,{\footnotesize$\pm$\,0.015} & 0.016\,{\footnotesize$\pm$\,0.824} & 0.016\,{\footnotesize$\pm$\,0.000} & 0.016\,{\footnotesize$\pm$\,0.007} \\
   & & 15k–30k & --- & \best{0.016\,{\footnotesize$\pm$\,0.000}} & \best{0.016\,{\footnotesize$\pm$\,0.041}} & \worst{2.34\,{\footnotesize$\pm$\,1.26}} & \best{0.016\,{\footnotesize$\pm$\,0.000}} & \best{0.016\,{\footnotesize$\pm$\,0.040}} \\
  \hline
  \multirow{24}{*}{\rotatebox{90}{General context-free}}
   & \multirow{3}{*}{S-Expression} & 0–5k & --- & --- & 0.016\,{\footnotesize$\pm$\,0.002} & 0.016\,{\footnotesize$\pm$\,0.002} & 0.016\,{\footnotesize$\pm$\,0.000} & 0.016\,{\footnotesize$\pm$\,0.000} \\
   & & 5k–15k & --- & --- & 0.016\,{\footnotesize$\pm$\,0.007} & 0.016\,{\footnotesize$\pm$\,0.920} & 0.016\,{\footnotesize$\pm$\,0.000} & 0.016\,{\footnotesize$\pm$\,0.000} \\
   & & 15k–30k & --- & --- & 0.047\,{\footnotesize$\pm$\,0.021} & \worst{2.30\,{\footnotesize$\pm$\,1.33}} & \best{0.016\,{\footnotesize$\pm$\,0.000}} & \best{0.016\,{\footnotesize$\pm$\,0.000}} \\
   & \multirow{3}{*}{TinyC} & 0–5k & --- & --- & 0.016\,{\footnotesize$\pm$\,0.006} & 0.016\,{\footnotesize$\pm$\,0.000} & 0.016\,{\footnotesize$\pm$\,0.000} & 0.016\,{\footnotesize$\pm$\,0.000} \\
   & & 5k–15k & --- & --- & 0.016\,{\footnotesize$\pm$\,0.000} & 0.016\,{\footnotesize$\pm$\,0.994} & 0.016\,{\footnotesize$\pm$\,0.000} & 0.016\,{\footnotesize$\pm$\,0.015} \\
   & & 15k–30k & --- & --- & \best{0.016\,{\footnotesize$\pm$\,0.000}} & \worst{2.41\,{\footnotesize$\pm$\,1.20}} & \best{0.016\,{\footnotesize$\pm$\,0.000}} & \best{0.016\,{\footnotesize$\pm$\,0.017}} \\
   & \multirow{1}{*}{Bool} & 0–5k & --- & --- & 0.016\,{\footnotesize$\pm$\,0.004} & 0.016\,{\footnotesize$\pm$\,0.000} & 0.016\,{\footnotesize$\pm$\,0.000} & 0.016\,{\footnotesize$\pm$\,0.000} \\
   & \multirow{1}{*}{Expr (ambig)} & 0–5k & --- & --- & 0.016\,{\footnotesize$\pm$\,0.531} & 0.016\,{\footnotesize$\pm$\,1.374} & 0.016\,{\footnotesize$\pm$\,1.195} & 0.016\,{\footnotesize$\pm$\,0.993} \\
   & \multirow{3}{*}{JSON (ambig)} & 0–5k & --- & --- & 0.016\,{\footnotesize$\pm$\,0.036} & 0.016\,{\footnotesize$\pm$\,0.101} & 0.016\,{\footnotesize$\pm$\,0.005} & 0.016\,{\footnotesize$\pm$\,0.002} \\
   & & 5k–15k & --- & --- & \worst{0.062\,{\footnotesize$\pm$\,0.041}} & \best{0.016\,{\footnotesize$\pm$\,0.055}} & \best{0.016\,{\footnotesize$\pm$\,0.096}} & \best{0.016\,{\footnotesize$\pm$\,0.002}} \\
   & & 15k–30k & --- & --- & 0.062\,{\footnotesize$\pm$\,0.333} & \worst{0.719\,{\footnotesize$\pm$\,1.448}} & \best{0.016\,{\footnotesize$\pm$\,0.462}} & \best{0.016\,{\footnotesize$\pm$\,0.487}} \\
   & \multirow{2}{*}{ANSI C} & 0–5k & --- & --- & \worst{0.031\,{\footnotesize$\pm$\,0.072}} & \best{0.016\,{\footnotesize$\pm$\,1.488}} & \best{0.016\,{\footnotesize$\pm$\,0.163}} & \best{0.016\,{\footnotesize$\pm$\,0.274}} \\
   & & 5k–15k & --- & --- & \best{0.094\,{\footnotesize$\pm$\,0.083}} & \worst{7.10\,{\footnotesize$\pm$\,4.56}} & \best{0.016\,{\footnotesize$\pm$\,1.734}} & \best{0.016\,{\footnotesize$\pm$\,1.659}} \\
   & \multirow{3}{*}{Pascal} & 0–5k & --- & --- & 0.016\,{\footnotesize$\pm$\,0.024} & 0.016\,{\footnotesize$\pm$\,0.893} & 0.016\,{\footnotesize$\pm$\,0.002} & 0.016\,{\footnotesize$\pm$\,0.000} \\
   & & 5k–15k & --- & --- & \best{0.016\,{\footnotesize$\pm$\,0.037}} & \worst{3.84\,{\footnotesize$\pm$\,1.96}} & \best{0.016\,{\footnotesize$\pm$\,0.003}} & \best{0.016\,{\footnotesize$\pm$\,0.020}} \\
   & & 15k–30k & --- & --- & 0.141\,{\footnotesize$\pm$\,0.064} & \worst{12.5\,{\footnotesize$\pm$\,1.7}} & 0.094\,{\footnotesize$\pm$\,0.035} & \best{0.062\,{\footnotesize$\pm$\,0.049}} \\
   & \multirow{2}{*}{Java} & 0–5k & --- & --- & 0.016\,{\footnotesize$\pm$\,0.061} & 0.016\,{\footnotesize$\pm$\,1.385} & 0.016\,{\footnotesize$\pm$\,0.000} & 0.016\,{\footnotesize$\pm$\,0.000} \\
   & & 5k–15k & --- & --- & \best{0.062\,{\footnotesize$\pm$\,0.135}} & \worst{4.95\,{\footnotesize$\pm$\,2.54}} & \best{0.016\,{\footnotesize$\pm$\,0.000}} & \best{0.016\,{\footnotesize$\pm$\,0.000}} \\
   & \multirow{2}{*}{C++} & 0–5k & --- & --- & 0.047\,{\footnotesize$\pm$\,0.239} & \worst{0.078\,{\footnotesize$\pm$\,5.131}} & \best{0.016\,{\footnotesize$\pm$\,0.561}} & \best{0.016\,{\footnotesize$\pm$\,0.580}} \\
   & & 5k–15k & --- & --- & 1.58 & \worst{11.6\,{\footnotesize$\pm$\,15.2}} & \best{0.016\,{\footnotesize$\pm$\,1.078}} & \best{0.016\,{\footnotesize$\pm$\,0.356}} \\
   & \multirow{1}{*}{CSS} & 0–5k & --- & --- & 0.016\,{\footnotesize$\pm$\,0.224} & 0.016\,{\footnotesize$\pm$\,0.046} & 0.016\,{\footnotesize$\pm$\,0.005} & 0.016\,{\footnotesize$\pm$\,0.002} \\
   & \multirow{1}{*}{HTML} & 0–5k & --- & --- & 0.016\,{\footnotesize$\pm$\,0.010} & 0.016\,{\footnotesize$\pm$\,1.191} & 0.016\,{\footnotesize$\pm$\,0.037} & 0.016\,{\footnotesize$\pm$\,0.000} \\
   & \multirow{1}{*}{Shell} & 0–5k & --- & --- & 0.016\,{\footnotesize$\pm$\,0.016} & 0.016\,{\footnotesize$\pm$\,0.295} & 0.016\,{\footnotesize$\pm$\,0.022} & 0.016\,{\footnotesize$\pm$\,0.020} \\
   & \multirow{1}{*}{SQL} & 0–5k & --- & --- & 0.016\,{\footnotesize$\pm$\,0.005} & 0.016\,{\footnotesize$\pm$\,0.000} & 0.016\,{\footnotesize$\pm$\,0.000} & 0.016\,{\footnotesize$\pm$\,0.000} \\
  \hline
\end{tabular}

  \vspace{-0.3cm}
  \label{tab:memoryBucket}
\end{table*}

\section{Results}\label{sec:results}
\Cref{tab:runtimeBucket} summarizes absolute median runtimes
across all 22 grammars, broken down by input-size bucket, the classification of the grammar, and the grammar itself.
Similarly \Cref{tab:memoryBucket} summarizes the peak memory usage while processing these inputs.

\subsection{RQ1. Comparison of Parser Families}

\noindent\textbf{Offline generalized parsers.}
\Cref{tab:cykValiant} shows our experimental results for offline generalized parsers.
Despite Valiant's superior asymptotic complexity of $O(\frac{n^3}{\log n})$ versus
CYK's $O(n^3)$, in practice, Valiant is dramatically slower than CYK across every tested grammar.
On Expr~(lr) at 20--100~tokens,
CYK takes 2.41~\millisecond while Valiant takes 796~\millisecond---a $330\times$ difference.
The gap widens further on ambiguous grammars:
Valiant reaches 3{,}306~\millisecond on Expr~(ambig) at 20--100~tokens, versus 2.26~\millisecond for CYK.
Our results show that these algorithms are inefficient.
Earley is faster than both CYK and Valiant on every grammar:
on Expr~(lr) at 20--100~tokens Earley completes in 0.19~\millisecond,
a $13\times$ speedup over CYK and a $4{,}000\times$ speedup over Valiant.
As these algorithms
are impractical even at token counts below 100, we do not evaluate them further.

\begin{result}
Both CYK and Valiant are impractical at realistic input sizes.
Earley outperforms both even on inputs of fewer than 100~tokens.
\end{result}

\noindent\textbf{Runtime of online generalized parsers.}
\Cref{tab:runtimeBucket} reports median runtime across input-size buckets
for Earley, GLL, RNGLR, and BRNGLR.
RNGLR and BRNGLR are the fastest generalized parsers by a consistent margin.
At 15k--30k~tokens on TinyPascal,
RNGLR takes 21~\millisecond versus 40~\millisecond for GLL and 157~\millisecond for Earley;
on S-Expr~LL-1 the same ordering holds at 26.4~\millisecond, 45.6~\millisecond, and 213~\millisecond.
Across all grammars and buckets where all four algorithms are applicable,
RNGLR and BRNGLR are typically two to three times faster than GLL
and four to ten times faster than Earley.
RNGLR and BRNGLR themselves are essentially indistinguishable,
typically within 15\% of each other in every case.

GLL's performance is sensitive to grammar structure.
On LL-friendly inputs it is competitive with RNGLR,
but it degrades severely on left-recursive grammars:
on JSON~(lr) at 15k--30k~tokens,
GLL takes 346~\millisecond while RNGLR takes 17~\millisecond---a $20\times$ gap.
On SQL, GLL reaches 0.08~\millisecond at 0-5k tokens 
versus 0.03--0.04~\millisecond for RNGLR and BRNGLR.
Earley is generally the slowest generalized parser.

\begin{result}
  RNGLR and BRNGLR are consistently the fastest
  generalized parsers,
  and are effectively equivalent to each other.
  GLL is competitive on LL-friendly grammars
  but degrades sharply on left-recursive inputs.
  Earley is consistently the slowest generalized parser.
\end{result}

\noindent\textbf{Memory consumption.}
\Cref{tab:memoryBucket} reveals a different ordering than runtime.
LL(1) and LR(1) parsers maintain negligible memory usage---at
or below the 0.016~MB measurement floor---across all grammars and input sizes.
RNGLR and BRNGLR are similarly memory-efficient,
remaining at 0.016~MB for the vast majority of grammars even at 15k--30k~tokens,
comparable in practice to the deterministic parsers.
Earley uses slightly more memory on a few grammars
(Pascal at 15k--30k: 0.141~MB; C++ at 5k--15k: 1.58~MB, S-Expr LL-1: 4.16~MB)
but remains modest in most cases.
\todo{RG: Earley and GLL performs oddly on S-Expr LL1}

GLL stands out as the only generalized parser with substantial memory growth for numerous grammars.
Its consumption rises steeply on larger and more complex grammars:
2.30--2.41~MB on S-Expression, TinyC, and TinyPascal at 15k--30k~tokens;
3.84~MB on Pascal at 5k--15k~tokens,
growing to 12.5~MB at 15k--30k~tokens;
7.10~MB on ANSI~C and 11.6~MB on C++ at 5k--15k~tokens.
RNGLR on the same grammars stays at 0.016~MB throughout.

\begin{result}
  RNGLR and BRNGLR are the most memory-efficient generalized parsers,
  matching deterministic parsers on most grammars.
\end{result}

\begin{figure}[tp]
  \centering
  \input{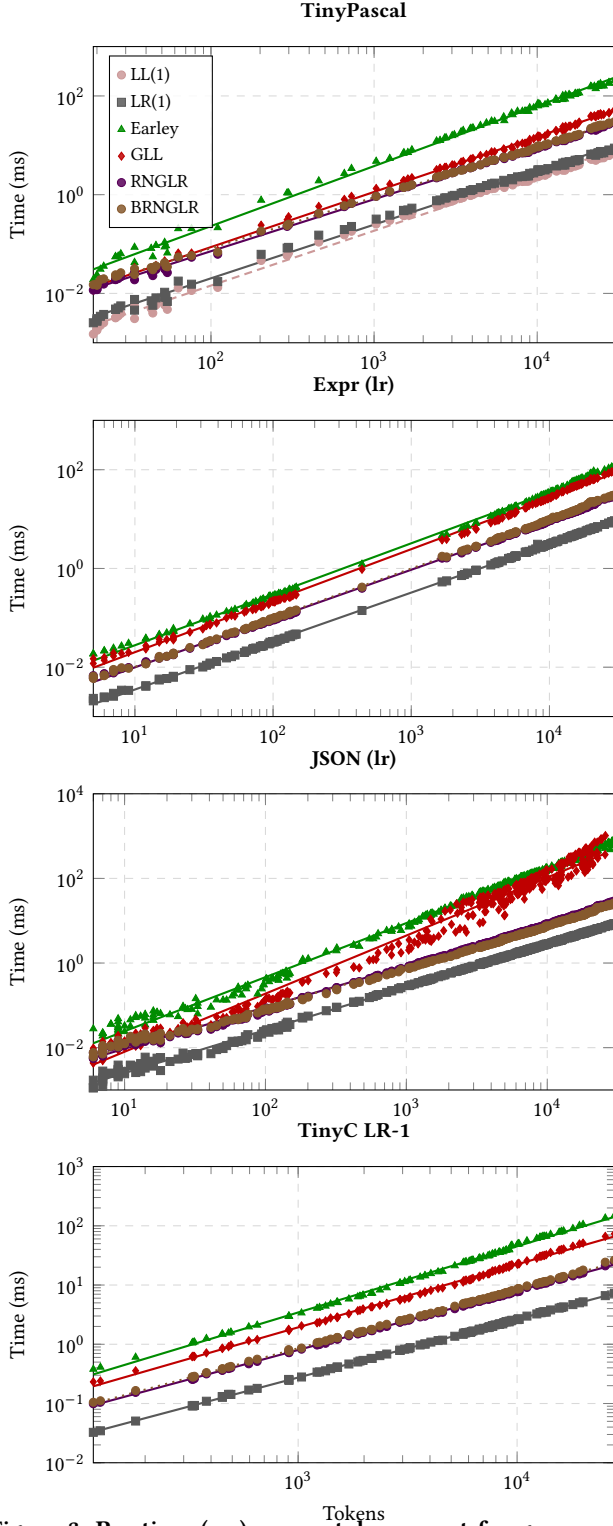}
  \vspace{-1cm}
  \caption{%
    Runtime (\millisecond) versus token count 
    for grammars with a known LR(1) baseline.}
  \label{fig:lrBaselineTime}
\end{figure}

\subsection{RQ2. Deterministic vs. Generalized Parsers}

\noindent\textbf{Runtime.}
\Cref{tab:runtimeBucket} and \Cref{fig:lrBaselineTime} compare deterministic and generalized parsers
across all input-size buckets.
LL(1) and LR(1) are the fastest parsers in every row where they are applicable,
typically running at 5--8~\millisecond at 15k--30k~tokens regardless of grammar,
while even the fastest generalized parser (RNGLR) runs at 15--26~\millisecond on most grammars
(up to 63--115~\millisecond on JSON variants) at the same input size.
LL(1) is consistently 10--20\% faster on LL(1) grammars.

Comparing on five LR(1) baseline grammars,
the median runtime overhead relative to LR(1) is:
RNGLR $\times 3.0$ (IQR: $\times 2.9$--$\times 3.1$, max $\times 4.6$);
BRNGLR $\times 3.0$ (IQR: $\times 2.7$--$\times 3.2$, max $\times 5.7$);
GLL $\times 6.0$ (IQR: $\times 4.5$--$\times 8.4$, max $\times 139$);
Earley $\times 10.0$ (IQR: $\times 7.9$--$\times 11.4$, max $\times 15.3$).

GLR's overhead is narrow and predictable:
the IQR spans only 0.2$\times$ around the median,
and no grammar class causes it to exceed $\times 5.7$.
GLL's overhead is considerably more variable:
its IQR is nearly twenty times wider than GLR's.

\begin{result}
  Deterministic parsers are 3$\times$ faster than the best generalized
  alternative at large input sizes.
  Among generalized parsers, GLR's overhead over LR(1)
  is narrow and predictable ($\times 3$, IQR $\pm 0.2$),
  while GLL's varies widely and can reach $\times 139$
  on left-recursive grammars.
\end{result}

\noindent\textbf{Memory.}
LL(1) and LR(1) stay at the 0.016~MB measurement floor across every grammar and
every input-size bucket in \Cref{tab:memoryBucket}.
Among generalized parsers,
RNGLR and BRNGLR also remain at 0.016~MB on most grammars,
matching the deterministic parsers in practice.
GLL diverges at larger inputs, reaching 2--4~MB on simpler grammars
and 12.5~MB on Pascal at 15k--30k~tokens.
Earley is more modest,
staying below 0.2~MB on most grammars
with the exception of C++ (1.58~MB at 5k--15k~tokens).

\begin{result}
  Deterministic parsers consume negligible memory regardless of input size.
  RNGLR and BRNGLR match this behavior on most grammars.
  GLL is the only parser whose memory grows substantially
  with input size and grammar complexity.
\end{result}

\subsection{RQ3. Effectiveness of Grammar Refactoring}

\begin{figure*}[tp]
  \centering
  \input{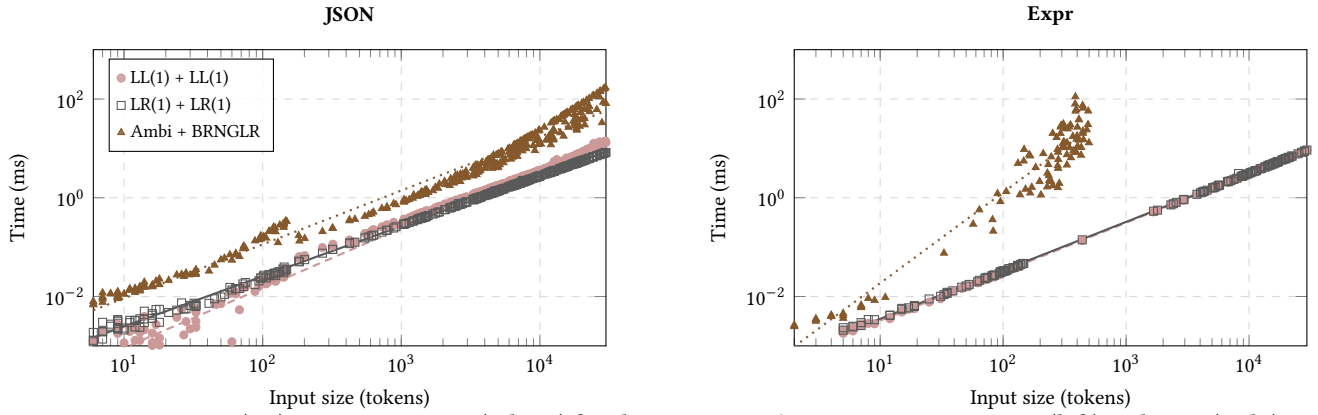}
  \vspace{-0.5cm}
  \caption{%
    Runtime (\millisecond) versus input size (tokens) for three grammar/parser pairings
    on JSON (left) and Expr (right).}
  \vspace{-0.0cm}
  \label{fig:grammarComparison}
\end{figure*}
\begin{figure*}[tp]
  \centering
  \vspace{0cm}
  \input{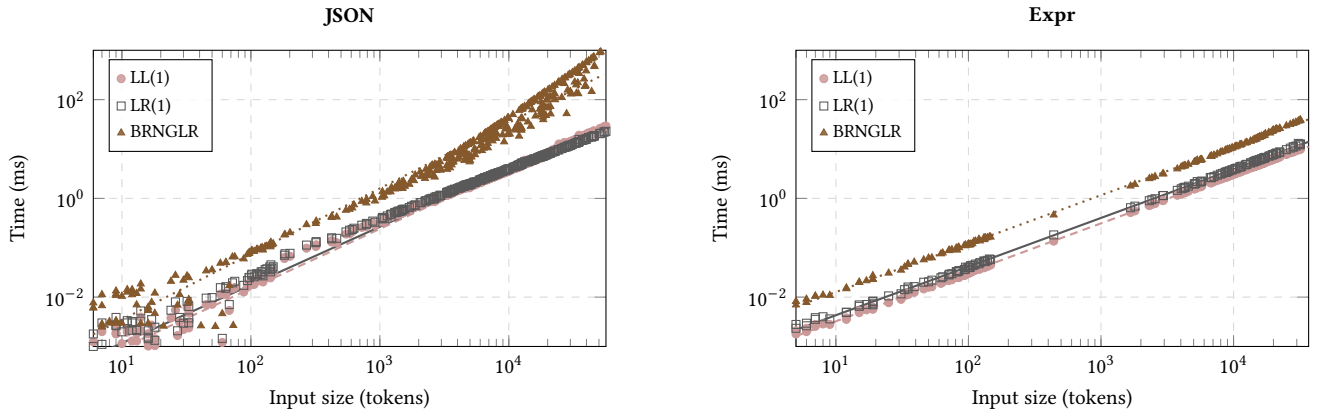}
  \vspace{-0.5cm}
  \caption{%
    Runtime (\millisecond) versus input size (tokens) for three parsers---LL(1),
    LR(1), and BRNGLR, same grammar.
    }
  \label{fig:ll1GrammarParserComparison}
\end{figure*}

We compare four variants of two representative grammars---JSON and Expr---across
all parsers to isolate the effect of grammar structure.
JSON and Expr are the only grammars in our benchmark
for which all four structural variants
(left-recursive, right-recursive, LL(1)-refactored, and ambiguous)
have been prepared;
the remaining grammars exist only in their natural form,
so a comparable cross-variant analysis is not possible for them.
The variants are:
a right-recursive form~(rr),
a left-recursive form~(lr),
an LL(1)-refactored form~(LL-1) that eliminates both recursion styles
in favor of iteration,
and an ambiguous form~(ambig) that retains the recursive structure
with underspecified associativity.

\noindent\textbf{Unlocking the deterministic tier.}
The primary payoff of grammar-hacking is that it enables LL(1) or LR(1) parsing.
As shown in \Cref{tab:runtimeBucket} and \Cref{fig:grammarComparison},
the LL(1)-refactored JSON grammar runs at 7.79~\millisecond (LL(1)) and 7.67~\millisecond (LR(1))
at 15k--30k~tokens.
The original right-recursive JSON grammar cannot be parsed by LL(1) at all,
and yields 342~\millisecond for Earley and 32.8~\millisecond for GLL---the best available
generalized parser for that variant.
Grammar-hacking to LL(1) thus represents a $4\times$ speedup
over the fastest generalized option for that grammar form.
The same pattern holds for Expr:
LL(1)-refactored Expr runs at 6.03~\millisecond with LL(1),
versus 43.1~\millisecond for GLL on the same grammar---a $7\times$ gap.

\begin{result}
  Grammar-hacking to LL(1)/LR(1) consistently reduces runtime by 4--7$\times$
  compared to the fastest generalized parser on the same grammar.
\end{result}

\noindent\textbf{Effect of grammar structure on generalized parsers.}
Grammar structure has a substantial and parser-specific impact 
even without a switch to deterministic parsing.

RNGLR performs best on left-recursive grammars.
On JSON~(lr) at 15k--30k~tokens, RNGLR takes 17.4~\millisecond;
on JSON~(rr) the same parser takes 63.6~\millisecond (a $3.7\times$ slowdown),
and on the LL(1)-refactored JSON~(LL-1) it takes 115~\millisecond (a $6.6\times$ slowdown).
The LL(1) refactoring---which replaces left-recursion with right-recursion to
satisfy LL(1) constraints---actually produces the worst-case grammar for RNGLR.

GLL's performance moves in the opposite direction.
It handles right-recursive and LL-refactored grammars well
(JSON~(rr): 32.8~\millisecond; JSON~LL-1: 46.9~\millisecond at 15k--30k~tokens),
but collapses on left-recursive input:
JSON~(lr) at 15k--30k~tokens takes 346~\millisecond for GLL,
a $10.5\times$ regression versus the right-recursive form.
Earley is comparatively grammar-structure-insensitive.
Across all four JSON variants at 15k--30k~tokens,
Earley ranges from 342~\millisecond~(rr) to 393~\millisecond~(lr)---a variation of only 15\%---
while GLL varies by a factor of 10 and RNGLR by a factor of seven
over the same grammar family.

Adding ambiguity to the recursive rules~(ambig) has negligible runtime impact
when the underlying recursion structure is already fixed.
JSON~(ambig) differs from JSON~(rr) by less than 1\% for all four generalized parsers.

\begin{result}
  Grammar structure has a large, parser-specific effect on generalized parsers.
  RNGLR benefits from left-recursion and degrades on right-recursion;
  GLL benefits from right and collapses on left-recursion.
  Earley is largely insensitive to recursion style.
  Refactoring a grammar to LL(1) to help GLL penalizes RNGLR.
\end{result}

\noindent\textbf{Memory across grammar variants.}
Memory consumption is also affected by grammar structure, most visibly for GLL.
On JSON~(lr) at 15k--30k~tokens, GLL stays at the 0.016~MB floor
despite taking 346~\millisecond to parse.
On JSON~(rr) at the same size, GLL grows to 0.828~MB;
on Expr~(LL-1) it reaches 2.42~MB.
RNGLR and BRNGLR remain at 0.016~MB across all JSON and Expr variants
at every input-size bucket.
GLL's memory and runtime costs are inversely coupled
across grammar variants: the grammar forms where it runs fastest
are those where it consumes the most memory.

\section{Discussion}\label{sec:discussion}

\noindent\textbf{GLR is the practical default for generalized parsing.}
Across every grammar and input size tested,
RNGLR and BRNGLR are the fastest generalized parsers by a consistent margin,
and simultaneously the most memory-efficient,
matching deterministic parsers at the measurement floor on most grammars.
The BRNGLR variant was motivated by a theoretical concern:
the standard RNGLR SPPF can grow to $O(n^3)$ nodes in the worst case,
whereas BRNGLR bounds this to $O(n^2)$ via binarization.
In our results,
RNGLR and BRNGLR typically differ by less than 15\% in runtime and are indistinguishable in memory,
suggesting that the pathological $O(n^3)$-node SPPF is rare in typical software engineering grammars.
For practitioners, the two algorithms are interchangeable;
RNGLR is simpler to implement and should be preferred. 

\noindent\textbf{GLL's memory and runtime costs are inversely coupled across grammar variants.}
GLL's performance profile is unstable across grammars.
On right-recursive and LL(1)-refactored grammars, GLL is fast---
JSON~(rr) at 15k--30k tokens takes 32.8~\millisecond---but it consumes significant memory,
reaching 0.828~MB on JSON~(rr) and 2.42~MB on Expr~(LL-1).
On JSON (lr), the pattern reverses:
JSON~(lr) at 15k--30k tokens takes 346~\millisecond, yet GLL remains at the 0.016~MB floor.
A deployment that selects GLL must carefully verify that GLL is indeed the fastest algorithm on such grammars.

\noindent\textbf{The cost of generality is modest for GLR.}
A 3$\times$ median runtime overhead relative to LR(1),
with an IQR of only 0.2$\times$,
means GLR's cost is both low and predictable:
for GLR, that cost is bounded and narrow across diverse real-world grammars.
GLL's overhead is less favorable---a median of 6$\times$ with an IQR nearly twenty times wider than GLR's,
and a worst-case of 139$\times$ on left-recursive JSON---
and Earley's median of 10$\times$ makes it the most expensive online option.
However, even for these parsers, the overhead is a constant factor on any given grammar;
the behavior is not inherently unbounded.
The practical implication is that GLR closes the performance gap with deterministic parsers
to a level that is acceptable for a wide range of software engineering applications.

\noindent\textbf{Grammar structure and parser choice interact.}
No single grammar form is universally optimal
once the algorithm family is fixed.
RNGLR is fastest on left-recursive grammars and degrades on right-recursive ones;
GLL is faster than RNGLR on the JSON LL(1)-refactored grammar, but collapses
on left-recursive ones.
This interaction has a concrete consequence: grammar-hacking a grammar to LL(1)---the
refactoring that introduces right-recursion and iteration to satisfy LL(1) constraints---improves
GLL by enabling faster parsing but simultaneously produces the worst-case grammar for RNGLR.
A practitioner who refactors a grammar expecting to improve performance across the board
may inadvertently degrade the parser they are actually using.
The recommendation is to choose a grammar refactoring in light of the target algorithm:
LR(1)-style refactoring (preserving or introducing left-recursion) suits the GLR family,
while LL(1)-style refactoring suits GLL.

\noindent\textbf{Grammar hacking is most valuable as a gateway to the deterministic tier.}
The primary benefit of grammar hacking is not marginal improvement within the generalized tier
but the ability to unlock LL(1) or LR(1) parsing entirely.
This 4--7$\times$ speedup is consistent, but
comes with maintenance costs:
refactored grammars are harder to read, 
and can be brittle under language evolution.
Grammar hacking should therefore be undertaken deliberately,
only when parsing is a demonstrated bottleneck.
For most software engineering tools,
the 3$\times$ GLR overhead is acceptable. 

\noindent\textbf{Practical guidance for SE tool builders.}
Our recommendation:
For tools that require the maximum parsing throughput and whose grammar can be kept strictly deterministic,
LL(1) or LR(1) remain the best choices.
For tools that prioritize grammar flexibility---accepting arbitrary context-free grammars without modification---RNGLR is the recommended default:
it offers the best combination of runtime and memory efficiency,
and its overhead relative to LR(1) is small and predictable.
GLL is a reasonable alternative when a top-down parsing style is architecturally desirable,
but practitioners must choose grammar forms carefully and be aware of memory growth.
Earley is the most conservative choice for correctness on exotic or highly ambiguous grammars,
but it is consistently the slowest online generalized parser and its use should be justified.
CYK and Valiant are not competitive for any practical input size in our benchmark
and should not be considered for production use.

\section{Threats to Validity}
\label{sec:threats}
\done{SR: could the trends in the results also be programming language dependent? or even programmer dependent? or hardware dependent? --- yes to all
(M1/ARM; vs x86). Implementation language (Rust) is already in Internal Validity.
Programmer dependency is partially mitigated by peer review.}
\done{SR: perhaps more importantly, is that the distribution of inputs could play an outsized role. e.g., changing how the fuzzer works/samples might lead to different trends? }

\noindent\textbf{Internal Validity.}
All parsers are implemented in Rust from scratch; implementation bugs may exist.
Shared data structures and grammar representations
reduce the surface area for implementation-specific bias.
Memory measurements rely on polling at 1\,\millisecond intervals,
which may miss short-lived peak allocations.
Timing is reported as the median over multiple iterations
to reduce the impact of transient system noise,
but background load on the experimental machine (Apple M1)
cannot be fully eliminated.

\noindent\textbf{External Validity.}
All benchmark inputs are generated by a grammar-based fuzzer.
Synthetic inputs produced by bounded-depth derivation
may not reflect the patterns
that appear in developer-written code.
Our experiments ran exclusively on an Apple M1;
processors differ in branch predictor, cache hierarchy, and instruction throughput
in ways that may affect relative rankings.
Our grammar corpus does not include real developer-written source files
or intentionally ambiguous grammars,
which are precisely the setting where generalized parsers have the most to offer.


\noindent\textbf{Construct Validity.}
Structural metrics (LL(1)/LR(1) class, production count, nonterminal count)
are used as proxies for grammar complexity,
but may not capture all aspects of runtime difficulty.

\section{Related Work}
\label{sec:relatedWork}
\noindent\textbf{General parsing as a strategy.}
Van den Brand et al.~\cite{vandenbrand1998current} identified friction in software renovation
caused by deterministic grammar constraints,
\done{the constraints are on deterministic grammars? --- deterministic grammar constraints is standard SE parsing phrasing;}
arguing that resolving LR conflicts manually imposes significant engineering cost.
They proposed GLR as a practical alternative.
Klint et al.~\cite{klint2005toward} formalised this concern
in the notion of \emph{grammarware engineering},
observing that grammar transformations required for deterministic parsing
obscure the intended language structure and increase long-term maintenance burden.

A practical benefit of generalized parsers is \emph{scannerless} parsing~\cite{salomon1989scannerless},
treating each character as a token without a separate lexer phase.
Van den Brand et al.~\cite{vandenbrand2002disambiguation} demonstrated that
Scannerless GLR (SGLR) is viable in industrial language processing tools,
and Economopoulos et al.~\cite{economopoulos2009faster} showed that
RNGLR can improve scannerless performance.
Our benchmark also operates in a scannerless mode,
allowing direct comparison without a lexer.

\todo{SR: in the following section, does it make sense to compare what earlier papers found with what this paper finds? --- TODO.}

\noindent\textbf{Existing empirical comparisons.}
The earliest cross-family comparison appears in Tomita's original GLR work~\cite{tomita1986efficient},
which argued that GLR substantially outperforms Earley
on natural-language grammars.
The original GLL paper~\cite{scott2010gll} noted that GLL parsers
perform well relative to RNGLR and BRNGLR, but provides no systematic evaluation.
\done{what are the 'grammar families'? the paper is mainly about algorithm families.}
Johnstone et al.~\cite{johnstone2006evaluating} provided the most detailed
pre-GLL empirical study of general parsers,
comparing three GLR variants (Farshi, RNGLR, BRNGLR) across
ANSI~C, Pascal, and COBOL grammars.
However, it predates GLL, omits Earley,
and evaluates each algorithm on a single input per grammar rather than a distribution of sizes.
Economopoulos's dissertation~\cite{economopoulos2006generalized}
provides a thorough theoretical treatment of RNGLR and BRNGLR,
with implementation details our work draws on,
but its empirical evaluation is similarly limited in grammar and input coverage.
McPeak and Necula's Elkhound~\cite{mcpeak2004elkhound} is a hybrid GLR parser
evaluated on construction speed rather than algorithmic throughput.
Afroozeh and Izmaylova~\cite{afroozeh2015faster} benchmarked an optimized GLL
against the original GLL implementation on Java, C\#, and OCaml grammars.
However, these results characterize within-family improvement only.
None of the above studies compares algorithms from all three major families---%
matrix-based, top-down online (Earley, GLL), and bottom-up online (GLR)---%
in a unified implementation across a grammar corpus of comparable breadth
and with statistical evaluation over inputs of varying length.

\noindent\textbf{Parsing algorithm improvements.}
Scott and Johnstone~\cite{scott2016structuring} introduced FGLL and RGLL,
structural refinements to GLL that reduce descriptor overhead.
Afroozeh and Izmaylova also proposed Iguana~\cite{afroozeh2015one},
a GLL-focused framework for practical scannerless parsing.
More recently, Scott and Johnstone~\cite{scott2026earley} adapted the Earley algorithm
to use a precomputed parse table in the style of GLR,
reporting performance approaching BRNGLR on their test grammars.
Our framework does not yet include this table-driven Earley variant;
it is a natural direction for future work.

\noindent\textbf{Other parsing approaches.}
Parr et al.~\cite{parr2014adaptive} introduced ALL(*),
which handles a broad class of grammars using adaptive lookahead.
ALL(*) does not handle indirect left recursion
and cannot enumerate all parse trees for ambiguous inputs,
placing it outside the general CFG class studied here.
Parsing Expression Grammars~\cite{ford2004parsing} use ordered choice
to enforce deterministic parsing by construction,
at the cost of potentially rejecting strings that belong to the intended language.
Parglare~\cite{dejanovic2022parglare} is a practical LR/GLR library for Python
that includes informal runtime comparisons between its LR and GLR modes,
though not as a controlled cross-algorithm benchmark.

\section{Data Availability}\label{sec:data}
Available at:
{\centering
\small
\url{https://doi.org/10.5281/zenodo.19231343}
}

\section{Conclusion}\label{sec:conclusion}
We presented the first unified, controlled benchmark
of six generalized parsing algorithms---CYK, Valiant, Earley, GLL, RNGLR, and BRNGLR---alongside
LL(1) and LR(1) baselines,
implemented in a single Rust framework with shared data structures.
Through an evaluation of 22 grammars, from simple expressions to complex languages such as C++ and Java,
we provide actionable guidance for software practitioners.
Our results show that the performance penalty
for choosing GLR is modest:
a median 3$\times$ overhead over LR(1). 
This is an acceptable margin for the broad class of tools---compilers, static analyzers, and language servers---that
currently rely on hand-written or grammar-hacked deterministic parsers.


Our work provides the empirical foundation
for principled adoption of generalized parsing as a safe and efficient default.

\bibliographystyle{ACM-Reference-Format}
\bibliography{references}

@String{CommunACM      = "Commun. {ACM}"}

@String{SIGPLANNot     = "{SIGPLAN} Not."}

@String{InfoControl    = "Inform. Control"}

@String{SciComputProg  = "Sci. Comput. Program."}

@String{ACMTransPLS    = "{ACM} Trans. Program. Lang. Syst."}

@String{ActaInform     = "Acta Inform."}

@String{TheorComputSci = "Theoret. Comput. Sci."}

@String{JComputSysSci  = "J. Comput. System Sci."}

@String{ComputJ        = "Comput. J."}

@String{NumerMath      = "Numer. Math."}

@String{ElecNotesTheor = "Electron. Notes Theor. Comput. Sci."}

@String{SLE            = "Proc. {ACM SIGPLAN} Int. Conf. Softw. Lang. Eng. ({SLE})"}

@String{CC             = "Proc. Int. Conf. Compiler Construction ({CC})"}

@String{IWPC           = "Proc. Int. Workshop Program Comprehension ({IWPC})"}

@String{Onward         = "Proc. {ACM} Symp. New Ideas, New Paradigms, Refl. Program. Softw. (Onward!)"}

@misc{brunsfeld2024tree,
  author       = {Max Brunsfeld and others},
  title        = {{Tree-sitter}},
  year         = {2024},
  howpublished = {Zenodo},
  note         = {Version 0.25.3},
  doi          = {10.5281/zenodo.4619183},
  url          = {https://github.com/tree-sitter/tree-sitter}
}

@article{ford2004parsing,
  title   = {Parsing Expression Grammars: A Recognition-based Syntactic Foundation},
  author  = {Ford, Bryan},
  journal = SIGPLANNot,
  volume  = {39},
  number  = {1},
  year    = {2004},
  month   = jan,
  pages   = {111--122},
  doi     = {10.1145/982962.964011}
}

@article{younger1967recognition,
  title   = {Recognition and Parsing of Context-Free Languages in Time $n^3$},
  author  = {Younger, Daniel H.},
  journal = InfoControl,
  volume  = {10},
  number  = {2},
  year    = {1967},
  month   = feb,
  pages   = {189--208},
  doi     = {10.1016/S0019-9958(67)80007-X}
}

@article{earley1970an,
  title   = {An Efficient Context-Free Parsing Algorithm},
  author  = {Earley, Jay},
  journal = CommunACM,
  volume  = {13},
  number  = {2},
  year    = {1970},
  month   = feb,
  pages   = {94--102},
  doi     = {10.1145/362007.362035}
}

@book{tomita1986efficient,
  title     = {Efficient Parsing for Natural Language},
  author    = {Tomita, Masaru},
  publisher = {Springer},
  address   = {Boston, MA},
  year      = {1986},
  doi       = {10.1007/978-1-4757-1885-0}
}

@article{johnstone2006evaluating,
  title   = {Evaluating {GLR} Parsing Algorithms},
  author  = {Johnstone, Adrian and Scott, Elizabeth and Economopoulos, Giorgios},
  journal = SciComputProg,
  volume  = {61},
  number  = {3},
  year    = {2006},
  month   = aug,
  pages   = {228--244},
  doi     = {10.1016/j.scico.2006.04.004}
}

@article{valiant1975general,
  title   = {General Context-Free Recognition in Less than Cubic Time},
  author  = {Valiant, Leslie G.},
  journal = JComputSysSci,
  volume  = {10},
  number  = {2},
  year    = {1975},
  month   = apr,
  pages   = {308--315},
  doi     = {10.1016/S0022-0000(75)80046-8}
}

@article{knuth1965on,
  title   = {On the Translation of Languages from Left to Right},
  author  = {Knuth, Donald E.},
  journal = InfoControl,
  volume  = {8},
  number  = {6},
  year    = {1965},
  month   = dec,
  pages   = {607--639},
  doi     = {10.1016/S0019-9958(65)90426-2}
}

@article{aycock2002practical,
  title   = {Practical {Earley} Parsing},
  author  = {Aycock, J.},
  journal = ComputJ,
  volume  = {45},
  number  = {6},
  year    = {2002},
  month   = jun,
  pages   = {620--630},
  doi     = {10.1093/comjnl/45.6.620}
}

@article{leo1991a,
  title   = {A General Context-Free Parsing Algorithm Running in Linear Time
             on Every {LR}($k$) Grammar without Using Lookahead},
  author  = {Leo, Joop M. I. M.},
  journal = TheorComputSci,
  volume  = {82},
  number  = {1},
  year    = {1991},
  month   = may,
  pages   = {165--176},
  doi     = {10.1016/0304-3975(91)90180-A}
}

@article{scott2006right,
  title   = {Right Nulled {GLR} Parsers},
  author  = {Scott, Elizabeth and Johnstone, Adrian},
  journal = ACMTransPLS,
  volume  = {28},
  number  = {4},
  year    = {2006},
  month   = jul,
  pages   = {577--618},
  doi     = {10.1145/1146809.1146810}
}

@article{scott2007brnglr,
  title   = {{BRNGLR}: A Cubic {Tomita}-style {GLR} Parsing Algorithm},
  author  = {Scott, Elizabeth and Johnstone, Adrian and Economopoulos, Rob},
  journal = ActaInform,
  volume  = {44},
  number  = {6},
  year    = {2007},
  month   = oct,
  pages   = {427--461},
  doi     = {10.1007/s00236-007-0054-z}
}

@article{scott2010gll,
  title   = {{GLL} Parsing},
  author  = {Scott, Elizabeth and Johnstone, Adrian},
  journal = ElecNotesTheor,
  volume  = {253},
  number  = {7},
  year    = {2010},
  month   = sep,
  pages   = {177--189},
  doi     = {10.1016/j.entcs.2010.08.041}
}

@inproceedings{johnstone2023a,
  title     = {A Reference {GLL} Implementation},
  author    = {Johnstone, Adrian},
  booktitle = SLE,
  year      = {2023},
  month     = oct,
  pages     = {43--55},
  doi       = {10.1145/3623476.3623521}
}

@phdthesis{economopoulos2006generalized,
  title  = {Generalised {LR} Parsing Algorithms},
  author = {Economopoulos, Giorgios Robert},
  school = {Royal Holloway, University of London},
  year   = {2006},
  month  = aug
}

@article{scott2026earley,
  title   = {Earley Table Traversing Parsers},
  author  = {Scott, Elizabeth and Johnstone, Adrian},
  journal = SciComputProg,
  volume  = {247},
  year    = {2026},
  month   = jan,
  pages   = {103335},
  doi     = {10.1016/j.scico.2025.103335}
}

@article{parr2014adaptive,
  title   = {Adaptive {LL(*)} Parsing: The Power of Dynamic Analysis},
  author  = {Parr, Terence and Harwell, Sam and Fisher, Kathleen},
  journal = SIGPLANNot,
  volume  = {49},
  number  = {10},
  year    = {2014},
  month   = oct,
  pages   = {579--598},
  doi     = {10.1145/2714064.2660202}
}

@inproceedings{mcpeak2004elkhound,
  title     = {Elkhound: A Fast, Practical {GLR} Parser Generator},
  author    = {McPeak, Scott and Necula, George C.},
  booktitle = CC,
  publisher = {Springer},
  year      = {2004},
  pages     = {73--88},
  doi       = {10.1007/978-3-540-24723-4_6}
}

@inproceedings{vandenbrand1998current,
  title     = {Current Parsing Techniques in Software Renovation Considered Harmful},
  author    = {van den Brand, Mark G. J. and Sellink, A. and Verhoef, C.},
  booktitle = IWPC,
  year      = {1998},
  month     = jun,
  pages     = {108}
}

@inproceedings{afroozeh2015one,
  title     = {One Parser to Rule Them All},
  author    = {Afroozeh, Ali and Izmaylova, Anastasia},
  booktitle = Onward,
  year      = {2015},
  month     = oct,
  pages     = {151--170},
  doi       = {10.1145/2814228.2814242}
}

@inproceedings{afroozeh2015faster,
  title     = {Faster, Practical {GLL} Parsing},
  author    = {Afroozeh, Ali and Izmaylova, Anastasia},
  booktitle = CC,
  publisher = {Springer},
  year      = {2015},
  pages     = {89--108},
  doi       = {10.1007/978-3-662-46663-6_5}
}

@article{strassen1969gaussian,
  title   = {Gaussian Elimination is not Optimal},
  author  = {Strassen, Volker},
  journal = NumerMath,
  volume  = {13},
  number  = {4},
  year    = {1969},
  month   = aug,
  pages   = {354--356},
  doi     = {10.1007/BF02165411}
}

@software{wiggers2019m4ri,
  author    = {Thom Wiggers and Renovate Bot},
  title     = {thomwiggers/m4ri-rust v0.3.2},
  year      = {2019},
  month     = aug,
  publisher = {Zenodo},
  version   = {v0.3.2},
  doi       = {10.5281/zenodo.3377514}
}

@misc{github2026search,
  title        = {Search Results for ``json parser'' Repositories},
  author       = {{GitHub}},
  year         = {2026},
  howpublished = {\url{https://github.com/search?q=json+parser&type=repositories}},
  note         = {Accessed: March 2026. Returns 29,800+ repositories}
}

@misc{langsec,
  title        = {Language-theoretic Security ({LangSec})},
  author       = {{LangSec Workshop}},
  year         = {2026},
  howpublished = {\url{https://langsec.org/}},
  note         = {Accessed: March 2026}
}

@misc{vandenbrand2026current,
  author = {Mark van den Brand},
  title = {{Current Parsing Techniques in Software Renovation Considered Harmful}},
  howpublished = {\url{https://www.cs.vu.nl/~x/ref/ref.html}},
  note = {Accessed: 2026-03-18},
}

@article{scott2016structuring,
    title={Structuring the GLL parsing algorithm for performance}, 
    volume={125}, ISSN={0167-6423}, DOI={10.1016/j.scico.2016.04.003}, 
    journal={Science of Computer Programming}, 
    author={Scott, Elizabeth and Johnstone, Adrian}, 
    year={2016}, 
    month=sep, 
    pages={1–22} 
}

@article{salomon1989scannerless,
    title={Scannerless NSLR(1) parsing of programming languages}, 
    volume={24}, ISSN={0362-1340}, DOI={10.1145/74818.74833}, 
    number={7}, 
    journal={SIGPLAN Not.}, 
    author={Salomon, D. J. and Cormack, G. V.}, 
    year={1989}, month=jun, pages={170–178} 
}

@inbook{vandenbrand2002disambiguation,
    address={Berlin, Heidelberg}, 
    series={Lecture Notes in Computer Science}, 
    title={Disambiguation Filters for Scannerless Generalized LR Parsers}, 
    volume={2304}, ISBN={978-3-540-43369-9}, 
    url={http://link.springer.com/10.1007/3-540-45937-5_12}, 
    DOI={10.1007/3-540-45937-5_12}, 
    booktitle={Compiler Construction},
    publisher={Springer Berlin Heidelberg}, 
    author={Van Den Brand, Mark G. J. and Scheerder, Jeroen and Vinju, Jurgen J. and Visser, Eelco}, 
    editor={Horspool, R. Nigel}, 
    year={2002}, 
    pages={143–158}, 
    collection={Lecture Notes in Computer Science} 
}

@inproceedings{economopoulos2009faster,
    address={Berlin, Heidelberg}, 
    title={Faster Scannerless GLR Parsing}, 
    ISBN={978-3-642-00722-4}, 
    DOI={10.1007/978-3-642-00722-4_10}, 
    booktitle={Compiler Construction}, 
    publisher={Springer}, 
    author={Economopoulos, Giorgios and Klint, Paul and Vinju, Jurgen}, 
    editor={de Moor, Oege and Schwartzbach, Michael I.}, 
    year={2009}, 
    pages={126–141}, 
    language={en} 
}

@inproceedings{lammel2001grammar,
  title={Grammar testing},
  author={L{\"a}mmel, Ralf},
  booktitle={International Conference on Fundamental Approaches to Software Engineering},
  pages={201--216},
  year={2001},
  organization={Springer}
}

@article{kirschner2022input,
  title={Input repair via synthesis and lightweight error feedback},
  author={Kirschner, Lukas and Soremekun, Ezekiel and Gopinath, Rahul and Zeller, Andreas},
  journal={arXiv preprint arXiv:2208.08235},
  year={2022}
}

@online{eaton2021parser,
  author  = {Phil Eaton},
  title   = {Parser generators vs. handwritten parsers: surveying major language implementations in 2021},
  year    = {2021},
  month   = aug,
  url     = {https://notes.eatonphil.com/parser-generators-vs-handwritten-parsers-survey-2021.html},
  urldate = {2026-03-22}
}

@online{eaton2021enumerating,
  author  = {Phil Eaton},
  title   = {Enumerating and analyzing 40+ non-V8 JavaScript implementations},
  year    = {2021},
  month   = sep,
  url     = {https://notes.eatonphil.com/javascript-implementations.html},
  urldate = {2026-03-22}
}

@online{tratt2011parsing,
  author  = {Laurence Tratt},
  title   = {Parsing: The Solved Problem That Isn't},
  year    = {2011},
  month   = mar,
  url     = {https://tratt.net/laurie/blog/2011/parsing_the_solved_problem_that_isnt.html},
  urldate = {2026-03-22}
}

@incollection{grune2008introduction,
  author    = {Dick Grune and Ceriel J. H. Jacobs},
  title     = {Introduction to Parsing},
  booktitle = {Parsing Techniques: A Practical Guide},
  publisher = {Springer New York},
  address   = {New York, NY},
  year      = {2008},
  pages     = {61--102},
  isbn      = {978-0-387-68954-8},
  url       = {https://doi.org/10.1007/978-0-387-68954-8_3}
}

@article{dejanovic2022parglare,
	author = {Igor Dejanovi{\'c}},
	doi = {https://doi.org/10.1016/j.scico.2021.102734},
	issn = {0167-6423},
	journal = {Science of Computer Programming},
	pages = {102734},
	title = {Parglare: A LR/GLR parser for Python},
	url = {https://www.sciencedirect.com/science/article/pii/S0167642321001271},
	volume = {214},
	year = {2022}}

@article{klint2005toward,
  title={Toward an engineering discipline for grammarware},
  author={Klint, Paul and L{\"a}mmel, Ralf and Verhoef, Chris},
  journal={ACM Transactions on Software Engineering and Methodology (TOSEM)},
  volume={14},
  number={3},
  pages={331--380},
  year={2005},
  publisher={ACM New York, NY, USA}
}

@article{arlazarov1970economical,
  author  = {Arlazarov, V. L. and Dinic, E. A. and
             Kronrod, M. A. and Faradjev, I. A.},
  title   = {On economical construction of the transitive closure
             of an oriented graph},
  journal = {Soviet Mathematics Doklady},
  volume  = {11},
  number  = {5},
  pages   = {1209--1210},
  year    = {1970},
}

@incollection{zeller2024fuzzing,
    author = {Andreas Zeller and Rahul Gopinath and Marcel B{\"o}hme and Gordon Fraser and Christian Holler},
    booktitle = {The Fuzzing Book},
    title = {Fuzzing with Grammars},
    year = {2024},
    publisher = {CISPA Helmholtz Center for Information Security},
    howpublished = {\url{https://www.fuzzingbook.org/html/Grammars.html}},
    note = {Retrieved 2024-06-30 18:31:28+02:00},
    url = {https://www.fuzzingbook.org/html/Grammars.html},
    urldate = {2024-06-30 18:31:28+02:00}
}

@online{kegler2023parsing,
  author  = {Jeffrey Kegler},
  title   = {Parsing: a Timeline},
  year    = {2023},
  month   = {July},
  url     = {https://jeffreykegler.github.io/personal/timeline_v3},
  note    = {Revision 9, 6 July 2023. Accessed: 2026-03-27}
}

\end{document}